\documentclass[pre,aps,showpacs,byrevtex,amsmath,amssymb,preprint]{revtex4}
\usepackage{graphicx}

\begin{document}

\author{Zuowei Wang}\email{wangzuo@mpip-mainz.mpg.de}
\author{Christian Holm}\email{holm@mpip-mainz.mpg.de}
\author{Hanns Walter M\"{u}ller} \email{hwm@mpip-mainz.mpg.de}
\affiliation{Max-Planck-Institut f\"ur Polymerforschung, Ackermannweg 10,
  D-55128 Mainz, Germany }
\title{A molecular dynamics study on the equilibrium magnetization
  properties and structure of ferrofluids} \date{\today}

\newcommand{\erfc}{{\mbox{erfc}}} \newcommand{\infd}{{\mbox{d}}}

\newcommand{\MM}{{\Bbb{M}}} \newcommand{\NN}{{\Bbb{N}}}
\newcommand{\RR}{{\Bbb{R}}} \newcommand{\ZZ}{{\Bbb{Z}}}

\newcommand{\CALP}{{\mathcal P}} \newcommand{\CALI}{{\mathcal I}}
\newcommand{\CALL}{{\mathcal L}} \newcommand{\CALM}{{\mathcal M}}
\newcommand{\CALQ}{{\mathcal Q}} \newcommand{\CALR}{{\mathcal R}}
\newcommand{\CALT}{{\mathcal T}}

\newcommand{\exa}{{^{\mbox{\footnotesize exa}}}}
\newcommand{\maxi}{{_{\mbox{\scriptsize max}}}}
\newcommand{\Mesh}{{_{\mbox{\scriptsize M}}}}
\newcommand{\nd}{{^{\mbox{\footnotesize nd}}}}
\newcommand{\opt}{{_{\mbox{\footnotesize opt}}}}
\newcommand{\self}{{^{\mbox{\scriptsize self}}}}

\newcommand{\D}{\displaystyle}
\newcommand{\VECE}{{\mathbf{E}}}
\newcommand{\VECk}{{\mathbf{k}}}
\newcommand{\VECI}{{\mathbf{I}}}
\newcommand{\VECv}{{\mathbf{v}}}
\newcommand{\VECl}{{\mathbf{l}}}
\newcommand{\VECm}{{\mathbf{m}}}
\newcommand{\VECM}{{\mathbf{M}}}
\newcommand{\VECn}{{\mathbf{n}}}
\newcommand{\VECr}{{\mathbf{r}}}
\newcommand{\VECz}{{\mathbf{z}}}
\newcommand{\VECH}{{\mathbf{H}}}
\newcommand{\VECF}{{\mathbf{F}}}
\newcommand{\VECR}{{\mathbf{R}}}
\newcommand{\VECmu}{{\mbox{\boldmath$\mu$}}}
\newcommand{\VECtau}{{\mbox{\boldmath$\tau$}}}
\newcommand{\VEComega}{{\mbox{\boldmath$\omega$}}}
\newcommand{\VECxi}{{\mbox{\boldmath$\xi$}}}
\newcommand{\VECchi}{{\mbox{\boldmath$\chi$}}}

\begin{abstract}
  We investigate in detail the initial susceptibility, magnetization curves,
  and microstructure of ferrofluids in various concentration and particle
  dipole moment ranges by means of molecular dynamics simulations. We use the
  Ewald summation for the long-range dipolar interactions, take explicitly
  into account the translational and rotational degrees of freedom, coupled to
  a Langevin thermostat. When the dipolar interaction energy is comparable
  with the thermal energy, the simulation results on the magnetization
  properties agree with the theoretical predictions very well.  For stronger
  dipolar couplings, however, we find systematic deviations from the
  theoretical curves. We analyze in detail the observed microstructure of the
  fluids under different conditions.  The formation of clusters is found to
  enhance the magnetization at weak fields and thus leads to a larger initial
  susceptibility. The influence of the particle aggregation is isolated by
  studying ferro-solids, which consist of magnetic dipoles frozen in at random
  locations but which are free to rotate. Due to the artificial suppression of
  clusters in ferro-solids the observed susceptibility is considerably lowered
  when compared to ferrofluids.
\end{abstract}

\pacs{PACS: 61.20.Ja Computer simulation of liquid structure;
75.05.Mm Magnetic fluids; 82.70.Dd Colloids; 82.20.Wt Computational 
modeling;simulation}

\maketitle

\section{Introduction}
%
Ferrofluids (dipolar magnetic fluids) are colloidal suspensions of
ferromagnetic particles of about $10nm$ diameter dispersed in a carrier liquid
\cite{rosensweig85a} that are usually stabilized against agglomeration by
coating particles with long-chain molecules (sterically) or decorating them
with charged groups (electrostatically). The small size of the particles
favors magnetic mono-domains with a magnetic moment proportional to the volume
of the magnetic grains. As a result, the particles interact with each other by
the long-range anisotropic dipole-dipole potential as well as the short-range
symmetric potentials, such as the steric repulsion, the electrostatic
repulsion and the van der Waals attraction. The study of both the
magnetization properties and structure of ferrofluids are of both fundamental
and application interests.

A ferrofluid system of sufficiently low concentration behaves like an ideal
paramagnetic gas. The interactions between the particles can be neglected, and
the physical properties of the system are well described by the one-particle
model \cite{shliomis74a}. In this case the equilibrium magnetization is
written in terms of the Langevin function
$\CALL(\alpha)=\coth(\alpha)-1/\alpha$
\begin{equation} \label{LangevinM}
M_L=M_{sat}\CALL(\alpha),\hspace{0.5cm} M_{sat}=\frac{N}{V}\frac{m}{\mu_{0}}.
\end{equation}
Here, $M_{sat}$ is the saturation magnetization of the fluid, $\alpha=m H/kT$
is the Langevin parameter, and $m$ is the magnetic moment, $N/V$ denotes the
number density of the particles, $H$ is the magnetic field, $T$ the
temperature, $k$ the Boltzmann constant, and $\mu_{0}=4\pi \times 10^{-7}H/m$.
Eq.(\ref{LangevinM}) leads to Curie's law for the initial susceptibility
$\chi_L$ which depends linearly on the particle concentration
\begin{equation} \label{Langsus}
\chi_L=\frac{N}{V}\frac{m^{2}}{3\mu_0 kT}.
\end{equation}
However, experiments on concentrated ferrofluids have revealed an essential
deviation from the Langevin formula \cite{pshenichnikov95a}. The
initial susceptibility increases with the particle concentration much faster
than that predicted in Eq.(\ref{Langsus}). This deviation is clearly due to
the particle interactions.

The influence of the particle interactions on the physical properties of
ferrofluids is investigated in terms of two dimensionless
parameters. One is the volume fraction of the particles
\begin{equation}
\phi=\frac{N}{V}\frac{\pi \sigma^3}{6},
\end{equation}
with $\sigma$ the particle diameter. The other is the dipolar coupling
constant
\begin{equation}
\lambda=\frac{m^2}{4\pi\mu_0 kT\sigma^3},
\end{equation}
which relates the  dipole-dipole interaction energy of two
contacting particles to the thermal energy $kT$. The two
parameters can be combined into the Langevin initial
susceptibility $\chi_L$ in Eq.(\ref{Langsus}) as
$\chi_L=8\phi\lambda$.

A number of theoretical models allow the evaluation of the equilibrium
magnetization of ferrofluids
\cite{onsager36a,grady83a,wertheim71a,morozov90a,buyevich92a,pshenichnikov96a,ivanov01a,huke00a}.
Among them, the mean-spherical model \cite{wertheim71a,morozov90a}, the
thermodynamic perturbation model \cite{buyevich92a} and the modified variant
of the effective field model \cite{pshenichnikov96a} have been shown to give
good results for the magnetic properties of ferrofluids with low or moderate
concentration of magnetic particles (up to $10-12\%$). Taking $\chi_L$ as a
universal parameter in weak magnetic fields, all the above approaches lead to
the same expression for the initial susceptibility
\begin{equation} \label{sus2ord}
\chi=\chi_L(1+\chi_L/3).
\end{equation}
For ferrofluids of higher concentration, larger particle size, or at low
temperature, Eq.(\ref{sus2ord}) is insufficient \cite{ivanov01a}.  Higher
order corrections in the dependence of $\chi$ or $M$ on $\phi$, $\lambda$, or
in general on $\chi_L$, turn out to be crucial.

An expression for $\chi$ up to cubic accuracy in $\chi_L$ can be obtained by
the mean-spherical model \cite{wertheim71a}, the Born-Mayer expansion method
\cite{huke00a} and the statistical model based on the pair correlation
function \cite{ivanov01a}, which reads as\cite{ivanov01a}
\begin{equation} \label{sus3ord}
\chi=\chi_L(1+\chi_L/3+\chi_L^2/144).
\end{equation}
The cubic term makes a noticeable difference of 5\% over
Eq.(\ref{sus2ord}) for $\chi_L \geq 4.14$.  The significance of the
third order contribution has been verified in comparison with
experimental measurements on the temperature dependence of $\chi$
\cite{ivanov01a}. However, in what we classify here as the strong
coupling limit, $\lambda>2$, the enhanced particle aggregation is
not sufficiently accounted for by the analytical studies, thus
giving rise to considerable deviations from the predictions of
Eqs.(\ref{sus2ord},\ref{sus3ord}). Here the magnetic properties
turn out to depend on $\Phi$ and $\lambda$ separately rather than
$\chi_L$ alone, a point which we will try to elucidate with our
simulation study.

Previous simulation works on dipolar (ferro) fluids were mainly
undertaken to explore the phase diagram. Large ranges of particle
concentration and dipolar coupling strength have been examined.
The results are found to depend substantially on the employed
short range interactions, such as hard sphere
\cite{weis93a,weis93b,levesque94a,camp00a}, soft sphere
\cite{wei92a,wei92b,stevens94a,stevens95a} and Lennard-Jones (LJ)
potentials \cite{leeuwen93a,stevens95b}. As an example, the
question whether a minimum amount of dispersive energy, i. e.,
the attractive van der Waals energy, is required to observe the
liquid-vapor coexistence is still under discussion
\cite{camp00a,leeuwen93a,stevens94a,stevens95a,stevens95b,levin99a,tlusty00a}.
Recent studies on the magnetic properties include the Monte Carlo
simulations on the dispersions of interacting super-paramagnetic
particles in a solid matrix \cite{chantrell00a} and on
ferrofluids with a finite spherical boundary
\cite{pshenichnikov00a}. In order to compare with the macroscopic
theories on ferrofluids, it is necessary to simulate the
ferrofluid systems with truly periodic boundary conditions.

In this paper, the Langevin dynamics simulation method is used to
study the equilibrium properties of ferrofluids. The long-range
dipolar interactions are dealt with the Ewald summation. The
initial susceptibility and the magnetization curves are
calculated as a function of $\phi$ and $\lambda$. The simulation
results are directly compared with the theoretical models. The
influence of particle aggregation is investigated systematically
by performing a cluster analysis of the microstructure of the
systems. The formation of aggregates is found to enhance the
magnetization of ferrofluids at weak magnetic fields. This is
confirmed by comparison with the calculated initial
susceptibility for a ferro-solid, i. e., a ferrofluid with the same
properties but with the particles' translational degree of freedom
frozen in at random positions.

The paper is organized as follows. We describe the simulation
method in Sec.II. The simulation results on the initial
susceptibility and the magnetization curves are given in Sec.III
(A) and (B), respectively. The cluster analysis of the
microstructure is presented in Sec.III (C). The initial
susceptibility of the ferro-solid systems is calculated in
Sec.III(D). We end with our conclusions in Sec.IV.

\section{Simulation Method}
The investigated ferrofluid systems consist of $N$ spherical
particles of diameter $\sigma$ distributed in a cubic simulation box
of side length $L$. Each particle has a permanent point dipole moment
${\bf m_i}$ at its center. Using periodic boundary conditions in all
spatial directions, the dipole-dipole interaction potential between
particle $i$ and $j$ is given by
\begin{equation} \label{Unormal}
U^{dip}_{ij} =\frac{1}{4\pi\mu_0}\sum_{\VECn \in \ZZ^3}
\bigg{\{} \frac{\VECm_i \cdot \VECm_j}{\mid \VECr_{ij}+\VECn L \mid^3} -
\frac{3[\VECm_i \cdot (\VECr_{ij}+\VECn L)][\VECm_j\cdot
(\VECr_{ij}+\VECn L)]}{\mid \VECr_{ij}+\VECn L\mid ^{5}} \bigg{\}},
\end{equation}
where $\VECr_{ij}=\VECr_i-\VECr_j$ is the displacement vector of
the two particles. The sum extends over all simple cubic lattice
points, $\VECn=(n_x,n_y,n_z)$ with $n_x$, $n_y$, $n_z$ integers.

In this work, we use the Ewald summation for dipolar systems to evaluate
Eq.(\ref{Unormal}) effectively, which gives
\cite{leeuw80a,allen87a,wang01a}
\begin{equation}
 U^{dip}_{ij}=U^{(r)}_{ij}+U^{(k)}_{ij}+U^{(self)}_{ij}+U^{(surf)}_{ij},
\end{equation}
where the real-space $U^{(r)}_{ij}$, the k-space
(reciprocal-space) $U^{(k)}_{ij}$, the self $U^{(self)}_{ij}$,
and the surface $U^{(surf)}_{ij}$ contributions are respectively
given by:
\begin{eqnarray}
U^{(r)}_{ij}&=&\frac{1}{4\pi\mu_0}\sum_{{\VECn} \in
  \ZZ^3}\bigg\{({\VECm}_i \cdot
{\VECm}_j) B(\mid {\VECr}_{ij}+{\VECn}\mid)- [{\VECm}_i \cdot
({\VECr}_{ij}+\VECn)][{\VECm}_j \cdot ({\VECr}_{ij}+\VECn)] C(\mid {\VECr}_{ij}
+{\VECn}\mid)\bigg\},\label{energyrr}\\
U^{(k)}_{ij}&=&\frac{1}{4\pi\mu_0 L^3}\sum_{\VECk \in \ZZ^3,{\VECk} \ne
  0}\frac{4\pi}{k^2}\exp[-(\pi k/\kappa L)^2] ({\VECm}_i \cdot {\VECk})({\VECm}_j \cdot {\VECk})
\exp(2 \pi i {\VECk} \cdot {\VECr}_{ij}/L),\label{energykk}\\
U^{(self)}_{ij}&=& -\frac{1}{4\pi\mu_0}\frac{2\kappa^3}{3\sqrt{\pi}}(m_i^2+m_j^2),\label{energyself}\\
U^{(surf)}_{ij}&=&\frac{1}{4\pi\mu_0}\frac{4\pi}{(2\mu_{BC}+1)L^3}{\VECm}_i\cdot {\VECm}_j.\label{energysurf}
\end{eqnarray}
with
\begin{eqnarray}
 B(r)&=&[\erfc(\kappa r)+(2\kappa r /\sqrt{\pi})\exp(-\kappa^2
 r^2)]/r^3,\label{functionB}\\
 C(r)&=&[3\erfc(\kappa r)+(2\kappa r/\sqrt{\pi})(3+2\kappa^2 r^2)
        \exp(-\kappa^2 r^2)]/r^5.\label{functionC}
\end{eqnarray}
Here $\erfc(x):=2\pi^{-1/2}\int_{x}^{\infty} \exp(-t^2) {\rm d} t$
denotes the complementary error function. The inverse length
$\kappa$ is the splitting parameter of the Ewald summation.
Eq.(\ref{energysurf}) assumes that the set of the periodic
replications of the simulation box tends in a spherical way
towards an infinite cluster and that the medium outside this
sphere is a uniform medium with magnetic permeability $\mu_{BC}$.
In this work, we use the metallic boundary condition
$\mu_{BC}=\infty$. The surface term vanishes in this case and
demagnetization effects do not occur. Thus the applied external
magnetic field $\VECH$ coincides exactly with the internal field.
The influence of different boundary conditions on the simulation
results will be the topic of a subsequent publication
\cite{wang02b}. The related formulas for the dipolar forces and
torques can be found in a previous paper \cite{wang01a}. The
theoretical estimates of the cutoff errors in the Ewald summation
derived there are used to determine the optimal values for the
Ewald parameters which enable us to minimize the overall
computational time at a predefined accuracy.

The short range interaction potential between the particles is
chosen to mimic the steric mechanism for stabilizing the
colloidal suspension. As mentioned in Sec.I, it is still not
settled which kind of potential is most suitable to model real
ferrofluids. It might actually depend on whether the solution is
electrically or sterically stabilized. In this work we adopt the
Lennard-Jones potential
\begin{equation} \label{LJpotential}
U^{LJ}_{ij} = 4\varepsilon \left[\left(\frac{\sigma}{r_{ij}}\right)^{12}
 - \left(\frac{\sigma}{r_{ij}}\right)^{6}-C(R_c)\right],
\end{equation}
where $C(R_c)=(\sigma/R_c)^{12}-(\sigma/R_c)^6$ with a cutoff radius of
$R_c=2^{1/6}\sigma$. In this way the particles have a purely repulsive
interaction force which smoothly decays to zero at $R_c$. This cutoff range is
smaller than the one usually used in the soft sphere potential which is
commonly adopted as $U_{SS}(r_{ij})=4\varepsilon (\sigma/r_{ij})^{12}$
\cite{wei92a,wei92b,stevens94a,stevens95a}, and thus closer to the
hard sphere potential as employed in most other theoretical calculations
\cite{ivanov01a,huke00a}. In this context the analytical calculations
of Ivanov \cite{ivanov01a} reveal that at least for the spatially
homogeneous systems the influence of different short range repulsive
potentials seems to be marginal.

The translational and rotational Langevin equations of motion of particle $i$
are given by \cite{allen87a,murashov00a}
\begin{eqnarray}
\CALM_{i} \dot{\VECv_{i}} &=&
\VECF_{i}-\Gamma_{T}\VECv_{i}+\VECxi_{i}^{T},
\label{LangevinT}\\
\VECI_i\cdot \dot{\VEComega_{i}} &=&
\VECtau_{i}-\Gamma_{R}\VEComega_{i}+\VECxi_{i}^{R},\label{LangevinR}
\end{eqnarray}
where $\CALM_i$ and $\VECI_i$ are the mass and inertia tensor of the particle,
$\Gamma_T$ and $\Gamma_R$ are the translational and rotational friction
constants, respectively. The Gaussian random force and torque have the
properties that the first moments vanish
\begin{eqnarray}
<\xi_{i\alpha}^T(t)> &=& 0, \\
<\xi_{i\alpha}^R(t)> &=& 0,
\end{eqnarray}
while the second moments satisfy
\begin{eqnarray}
<\VECxi_{i\alpha}^T(t)\cdot \VECxi_{j\beta}^T (t^{\prime})>=
6kT\Gamma_T\delta_{ij}\delta_{\alpha\beta}\delta(t-t^{\prime}),\\
<\VECxi_{i\alpha}^R(t)\cdot \VECxi_{j\beta}^R(t^{\prime})>=
6kT\Gamma_R\delta_{ij}\delta_{\alpha\beta}\delta(t-t^{\prime}),
\end{eqnarray}
where $\alpha$ and $\beta$ denote the $x,y,z$ components in the Cartesian
coordinates, respectively. Introducing the dipolar and short range terms into
Eqs.(\ref{LangevinT}) and (\ref{LangevinR}), the dimensionless equations of
motion can now be written as:
\begin{eqnarray}
\dot{\VECv^{\ast}_i}=\sum_{j\ne i}(\VECF_{ij}^{dip \ast}+
\VECF_{ij}^{LJ \ast}) - \Gamma_{T}^{\ast}
\VECv_{i}^{\ast}+\VECxi_{i}^{T\ast},\label{LangevinTd} \\
\VECI_i^{\ast}\cdot \dot{\VEComega_{i}^{\ast}}=\sum_{j\ne i}
\VECtau_{ij}^{dip \ast}+\VECm^{\ast}_{i}\times\VECH^{\ast}
-\Gamma_{R}^{\ast}\VEComega_{i}^{\ast}+\VECxi_{i}^{R\ast},
\label{LangevinRd}
\end{eqnarray}
where the variables are given in dimensionless form reduced by
the following units: length $r^{\ast}=r/\sigma$, dipole moment
$m^{\ast 2}=m^2/4\pi\mu_0 \epsilon\sigma^3$, moment of inertia
$I^{\ast}=I/(\CALM \sigma^2)$, time
$t^{\ast}=t(\epsilon/\CALM\sigma^2)^{1/2}$, the friction constants
$\Gamma_T^{\ast}=\Gamma_T (\sigma^2/\CALM\epsilon)^{1/2}$ and
$\Gamma_R^{\ast}=\Gamma_R/(\CALM\sigma^2\epsilon)^{1/2}$,
magnetic field $H^{\ast}=H(4\pi\mu_0\sigma^3/\epsilon)^{1/2}$ as
well as temperature $T^{\ast}=kT/\epsilon$. The values of the
dimensionless friction constants do not affect the equilibrium
properties. Here we adopt $\Gamma_T^{\ast}=10.0$ and
$\Gamma_R^{\ast}=3.0$, because in our simulations these values
had been found to speed up the equilibration time. These values
are also close to that used in Ref.\cite{murashov00a}. A value of
$I^{\ast}=0.4$ is used for the dimensionless moment of inertia,
corresponding to that of a rigid sphere with diameter $\sigma$.

The simulations were performed at constant temperature
$T^{\ast}=1$. The orientational coordinates of the particles were
expressed in terms of quaternion parameters and the equations of
motions were integrated with a leap-frog algorithm
\cite{allen87a}.  A reduced time step $\Delta t^{\ast}=0.002$ was
employed in all simulations. The runs were started from initial
configurations with random particle positions and dipole moment
orientations. For each case, the system was at first equilibrated
for a period of 50 000 time steps. The magnetization and
structural properties were then calculated from the data for
another period of at least 200 000 time steps. Error bars for the
simulation results were determined by dividing the simulation
runs into blocks of 10 000 time steps and calculating an estimate
for the standard deviation of the mean \cite{kusalik90a}.

\section{Results and Discussion}
%
Results were obtained from simulations on systems with $N=1000$ particles. The
number density of the particles was mostly taken in the range of
$\rho^{\ast}=N/V^{\ast}\leq 0.45$, corresponding to volume fractions
$\phi=\rho^{\ast}\pi/6 \leq 0.236$. This covers the typical particle
concentration ranges which are obtainable in real ferrofluids. A few more runs
were also carried out at higher concentrations in case of smaller $\lambda$ in
order to reach a sufficient large $\chi_L$. The values of the dipolar coupling
constant $\lambda$ were mainly chosen to be $1,2,3$ and $4$. To get an
estimate of realistic $\lambda$-values, we consider a ferrofluid sample
consisting of monodispersed magnetite particles. The bulk magnetization of
this material is $M_d=4.46\times 10^5$ A/m. With a magnetic core diameter of
the particles of $\sigma=10$ nm, the value of $\lambda$ amounts to about $1.3$
at room temperature $T=300K$. For larger particles with $\sigma=13$ nm,
$\lambda$ rises up to $2.9$ under the same conditions. Note that $\lambda$ may
further be enlarged by decreasing the temperature or using cobalt as working
material ($M_d=14.01\times 10^5$ A/m), for which the previous $\lambda$-values
increase by a factor of around $3.14$. We therefore also carried out some
simulations for $\lambda=8$.

In all simulations, we fixed the root mean square (rms)
absolute errors in the dipolar forces to
$\Delta F^{dip}\le 10^{-4}m^2/4\pi\mu_0\sigma^4$,
which corresponds to $10^{-4}/6$ of the attractive force between two contacting
particles with dipole moments in parallel alignment. The optimal
values of the Ewald parameters had been determined separately for
each system \cite{wang01a}. As an example, for a sample with
$\rho^{\ast}=0.1$ we get $r_{c}=0.425L$, $\kappa=6.78/L$ and
$K_c=6\times 2\pi/L$. While $\rho^{\ast}=0.3$ yields $r_{c}=0.46L$,
$\kappa=6.84/L$ and $K_c=7\times 2\pi/L$ to be the optimum values,
respectively.

\subsection{Initial susceptibility}
%
The influence of the particle interactions on the magnetic
properties is most evident at weak fields, so we start with the
investigation of the initial susceptibility $\chi$. The value of
$\chi$ is determined from the linear relationship of
$\VECM=\chi\VECH$ at $H \rightarrow 0$. Simulations are thus
performed to get the $M(H)$ curves at weak fields. The equilibrium
magnetization $\VECM$ is evaluated by the prescription
\begin{equation}
\VECM=\frac{1}{\tau}\int^{t_0+\tau}_{t_0}
(\frac{1}{\mu_0 V}\sum_{i=1}^{N}\VECm_{i}) dt,
\end{equation}
where $t_0$ is the time needed for the system to establish
the equilibrium state.

To study the weak field magnetization, we increase the Langevin parameter
$\alpha$ from $0$ to $0.5$. As expected, the calculation of $M$ at $\alpha=0$
does not show the existence of a spontaneous magnetization, because $\phi$ is
still not high enough. Other works \cite{stevens95a} have shown that
the transition to a ferromagnetic state occurs at $\phi \approx 0.44$ for
$\lambda=4$ .  For higher values of $\lambda$, the critical concentration is
around $\phi\approx 0.35$ for $\lambda=6.25$ \cite{weis93b,stevens95a}
and $\phi \approx 0.31$ for $\lambda=9$ \cite{wei92b}. These
conditions are clearly not encountered in our simulations.

The dimensionless magnetization
$M^{\ast}(=M/\sqrt{4\pi\epsilon/\mu_0\sigma^3})$ of the system
with $\lambda=3$ is shown in Fig.1 as a function of $\alpha$ and
$\phi$. The results for other values of $\lambda$ display a
similar behavior. It can be seen that $M^{\ast}$ at first
increases linearly with $\alpha$, then at higher field values
sub-linearly. We introduce $\alpha_{lin}$ as a definition for the
maximum value of $\alpha$ where $M^{\ast}$ still can be
considered to obey the linear relationship. Evidently, at a fixed
value of $\lambda$  this $\alpha_{lin}$-value decreases with
growing $\phi$. The data in Fig.1 show that when $\phi$ is
increased from $0.052$ to $0.209$, $\alpha_{lin}$ decreases from
$0.45$ to about $0.25$. On the other hand, at a fixed particle
concentration, rising $\lambda$ also implies a reduction of
$\alpha_{lin}$. This demonstrates that the range over which $M$
is proportional to $H$ (or equivalently $\alpha$) shrinks with
the increase of $\chi_L$. For a quantitative evaluation of the
initial susceptibility we need to determine the linear region for
each case separately. This is accomplished with the help of a
linear regression fit with error weighting \cite{press92a}. The
basic idea of that method is to minimize the so-called
$\xi$-square merit function
\begin{equation}\label{regression}
\xi^2=\sum^{N_d}_{i=1}(\frac{M_i- M(0) -\chi H_i}{\Delta M_i})^2
\end{equation}
so as to get $\chi$ and its error bar. Here $\Delta M_i$ is the error bar
related to the simulation data of $M_i$. In Fig.1 the data were taken with a
step size of $\Delta \alpha=0.05$ giving $11$ data points on each curve. Since
no spontaneous magnetization occurs, the initial data point is always at
$M(0)=0.0$. For each curve, a series of linear fittings was generated by
increasing $N_{d}$ in Eq.(\ref{regression}) from $5$ to $11$. The fit which
gives the minimum value of $\xi^2$ per degree of freedom is taken to be the
optimum linear fitting to the data.  The associated $N_{d}$-value is referred
as $N_{d,opt}$. The slope $\chi$ and its error bar $\Delta \chi$ determine the
initial susceptibility. The value of $\alpha_{lin}$ is roughly given by the
product of $\Delta \alpha$ with $N_{d,opt}-1$.
The results of this data fit including the error estimates are
compiled in Table.I. The degree of reliability of the results is
reflected by the high goodness-probability $Q$ \cite{press92a}.
The resulting straight lines $M=\chi H$, or in dimensionless form
$M^{\ast}=\chi\alpha/4\pi\sqrt{\lambda/T^{\ast}}$, are also
displayed in Fig.1 to show how they fit to the simulation data.

Fig.2(a) shows the calculated initial susceptibility $\chi$ as a
function of $\lambda$ and $\phi$, along with the analytical
results from Eqs.(\ref{sus2ord}) and (\ref{sus3ord}). Since the
Langevin susceptibility $\chi_L$ has been taken as the universal
parameter in these theoretical predictions
\cite{buyevich92a,pshenichnikov96a,ivanov01a}, we also map the
data of $\chi$ on $\chi_L$ in Fig.2(b) to show the scaling
behavior more directly. In addition to the results in Fig.2(a),
some more data was produced at higher concentrations (up to
$\phi=0.419$) for the case of $\lambda=1$ in order to get a
larger range of $\chi_L$.  We did not combine such high
concentrations with larger values of $\lambda$, since the close
proximity to the magnetic phase separation would make the
discussion more complex, rendering the results more sensitive to
the specific short-range interaction.

For the weak interaction range $\lambda \le 2$ we find perfect
agreement with Eqs.(\ref{sus2ord}) and (\ref{sus3ord}) for all
investigated volume fractions. Actually, because for these
coupling parameters we can reach only up to $\chi_L <4$, both
analytical predictions yield equally good results for our data
and are indistinguishable within the error bars. For $\lambda =3$
we see small but significant deviations at low values of $\phi$,
meaning that already at low values of $\chi_L$ neither
Eqs.(\ref{sus2ord}) nor (\ref{sus3ord}) are accurate. However, at
high volume fractions ($\chi_L > 3$), the third order equation
(\ref{sus3ord}) is compatible within error bars. For the
numerical values of the data see Table~1, where also the values
of the analytical results for different orders are given for
comparison. Even more drastic deviations from the theoretical
predictions are found for $\lambda = 4$ at low concentrations, in
the range $\chi_L \leq 6$.  We observe again the trend, that at
higher volume fractions the third order equation (\ref{sus3ord})
becomes compatible with our data. We can conclude that the
parameter $\lambda$, rather than $\chi_L$, becomes the dominating
control parameter for $\lambda>2$. This effect can be attributed
directly to the microstructure formation in the system. When
$\lambda \leq 2$, what we consider to be the weak interaction
limit, the dipolar potential is comparable to the thermal
fluctuation. Then the interparticle correlations are weak, and
the present theoretical models work well. However, going to
stronger interactions with $\lambda>2$, leads to a considerable
amount of particle aggregation. The resulting clusters amplify
the sensitivity of the system to weak external magnetic fields
and thus lead to an enhanced magnetic susceptibility. We will
come back to this point in Sec. \ref{subsec:micro}.

The influence of the volume fraction seems to be weak in the investigated
parameter regime. Noteworthy is that the third order equation seems to
describe the available data well at larger volume fraction and high coupling
parameter $\lambda$, even better than it does at intermediate values. On the
basis of our available data we cannot decide if this finding is a mere
coincidence. To our knowledge there is no theoretical reason to believe that
the third order equation should become reliable again at these high values of
$\chi_L$, where higher order corrections of the form $\phi^m \lambda^n$ with
$m$, $n$ arbitrary, are expected to become important.  Corrections linear in
$\phi$, but up to ninth order in $\lambda$ have been assembled for the
Born-Mayer expansion method in Ref.  \onlinecite{huke00a}. These additional
terms are all positive and therefore necessarily increase the susceptibility.
We have found that they improve the theoretical prediction of $\chi$ only at
very small concentrations ($\phi < 0.05$) for $\lambda>2$, but they
dramatically overestimate the simulation results at larger values of $\phi$.

Note that the simulations presented in
Ref.\onlinecite{pshenichnikov00a} give rise to a susceptibility
$\chi$ that {\em decreases} with growing $\lambda$ if $\chi_L$ is
fixed, which is {\it opposite} to our findings in Fig.2(b). We
suspect that this trend may be induced by finite size effects,
since a {\em finite} spherical simulation volume with only a few
hundred particles had been used in their simulations (without a
periodic replication to properly account for the long range
dipolar interaction as is done in the present work). We shall
clarify this point in a subsequent publication\cite{wang02b}.

\subsection{Magnetization curves}

In this section, we extend the calculation of $M$ to a larger
range of the applied external field strength $\alpha$ in order to
get the full magnetization curves. For moderately concentrated
ferrofluids, the `modified mean-field model'
\cite{pshenichnikov96a} describes the magnetization curve in
terms of the Langevin magnetization $M_L$ [Eq.(\ref{LangevinM})]
as follows
\begin{equation} \label{magzthe}
M(H)=M_L(H_e),
\end{equation}
where the effective field acting on an individual ferro-particle is
given by
\begin{equation} \label{magz2ord}
H_e=H+M_L(H)/3,
\end{equation}
rather than $H_e = H + M/3$ as done in the Weiss model. The
resulting expression for the initial susceptibility is
Eq.(\ref{sus2ord}). A higher order description is obtained by
replacing $H_e$ in Eq.(\ref{magz2ord}) by
\begin{equation}\label{magz3ord}
H_e=H+\frac{M_L(H)}{3}+\frac{1}{144}M_L(H)\frac{dM_L(H)}{dH},
\end{equation}
which in the weak field limit reduces to Eq.(\ref{sus3ord}) for $\chi$
\cite{ivanov01a}. The effective Langevin parameters corresponding
to Eqs.(\ref{magz2ord}) and (\ref{magz3ord}) can be written as
$\alpha_e=\alpha+\chi_L\CALL(\alpha)$ and $\alpha_e=\alpha+
\chi_L\CALL(\alpha)+\frac{\chi_L^2}{16}\CALL(\alpha)\frac{d\CALL(\alpha)}
{d\alpha}$, respectively. This reflects the $\chi_L$-dependence of
these two theoretical predictions. The Born-Mayer expansion
described $\chi$ in a different way, where its dependence on $\phi$
and $\lambda$ are given separately \cite{huke00a}.

We investigated the magnetization curves for two different cases. In the
first case $\chi_L$ is fixed to a small value of $1.256$ where the predictions
using Eqs.(\ref{magz2ord}) and (\ref{magz3ord}) practically coincide. Then
$\lambda$ is taken to be $1,2,4$ and $8$, respectively. To keep $\chi_L$
constant, the volume fraction $\phi$ had to be adapted from $0.157$ down to
$0.0196$. In the second case, we take a larger value of
$\chi_L=5.026$ which makes the difference between the two
theoretical predictions distinguishable by a few percentage.
Only $\lambda=3$ and $4$ are discussed. The
corresponding volume fractions are $\phi=0.209$ and $0.157$,
respectively. In the latter case, we
did not include the results for $\lambda \leq 2$ due to the limitation of the
particle concentrations and for $\lambda=8$ due to the extremely slow dynamics
which prohibited us from reaching  equilibrium distributions
within our computational limits.

The calculated magnetization curves are shown in Fig.3 together
with the theoretical predictions using
Eqs.(\ref{magzthe}-\ref{magz3ord}). For a more explicit
demonstration of the particle interaction the inlays display the
difference between the simulation data and the Langevin model
$(M-M_L)/M_{sat}$. The characteristic maximum of this difference
is found to be at intermediate field strength of $\alpha \approx
1$, where it may reach more than $35\%$.  The results in Fig.3(a)
reveal again the essential role of the dipolar coupling constant
$\lambda$. When $\lambda$ equals $1$ and $2$, the simulation and
theoretical curves agree with each other very well.  But for
larger $\lambda$ the calculated magnetization $M$ rises with
$\alpha$ much faster than the theoretical predictions, especially
in the weak field regime. This reflects the same result already
observed in Fig.2, that $\chi$ increases with $\lambda$ at low
and moderate concentrations if $\chi_L$ is held fixed. Due to
this effect, the peak positions of the increment $(M-M_L)/M_{sat}$
shift to smaller values of $\alpha$ with the increase of
$\lambda$. In Fig.3(b), the volume fraction of the particles are
relatively high. The maximum absolute difference between the two
analytical predictions occurs at $\alpha=0.62$ where the result
given by using Eq.(\ref{magz3ord}) is about $4.3\%$ larger than
that by Eq.(\ref{magz2ord}). Our simulation results are found to
be consistent with both of them within the range of error bar. A
quantitative analysis of the data indicates that
Eq.(\ref{magz3ord}) gives a better description in the weak field
regime. This consequently leads to a better agreement between the
simulation results on $\chi$ with the prediction of
Eq.(\ref{sus3ord}). However, both data sets for $\lambda=3$ and
$\lambda=4$ show deviations from the theoretical curves at larger
values of $\alpha$ which cannot be explained by statistical
errors. Also here we infer that higher order corrections are
needed to more accurately describe the magnetization curves.

\subsection{Microstructure analysis}\label{subsec:micro}

The microstructure and its relation to the magnetic properties of ferrofluids
are investigated by performing a cluster analysis. In conventional cluster
analysis, agglomerates are usually defined on the base of the spatial
proximity between the particles or by means of an energy criterion. For
ferrofluids the latter one is more favorable, because the anisotropic dipolar
interaction implies that two neighboring particles can form a stable bonding
only if their dipole moments are roughly aligned in head-to-tail orientation.

The energetically based cluster analysis of ferrofluids has been done in
several different ways
\cite{weis93a,levesque94a,stevens95a,pshenichnikov00a,coverdale98a}.  In this
work, we adopt the definition of Refs.  \onlinecite{weis93a,levesque94a} and
\onlinecite{stevens95a} where two particles are considered to be bound if
their dipolar potential energy is less than a predetermined value $U_{bond}$.
The influence of using different threshold values for $U_{bond}$ has been
investigated in a recent work \cite{tavares99a}. The values of $-1.4 \lambda
kT$ \cite{weis93a,levesque94a} and $-1.5\lambda kT$ \cite{stevens95a} were
shown both to give good results. In our calculations we have varied $U_{bond}$
from $-1.4\lambda kT$ to $-1.6\lambda kT$. Only slight quantitative changes of
the results have been found without a qualitative difference. The results
shown here have been evaluated for $U_{bond}=-1.5 \lambda kT$, i. e., $75\%$
of the contact energy of two perfectly co-aligned dipolar particles. The above
cluster definition implies that aggregates are mainly linear chains and rings.
Thicker particle agglomerates or branched chains, however, occur very rarely
in our simulations.

The size of a cluster is defined as the number of particles belonging to it.
We first set up the connectivity matrix by means of our cluster definition,
and then use it to evaluate the sizes and numbers of clusters in a given
configuration. The average cluster size is defined by
\begin{equation}
S_{avg}=<\sum_{s} s n_s/\sum_s n_s>.
\end{equation}
where $n_s$ is the number of clusters having size $s$, and the
triangular brackets denote the time average, or equivalently, the
average over the configuration space.

Fig.4 plots $S_{avg}$ as a function of $\lambda$ and $\phi$ for the case of
zero field ($\alpha=0$). Clearly, increasing the dipolar coupling $\lambda$
leads to a larger cluster size $S_{avg}$. However, in the weak coupling limit
$\lambda=1$ or $2$, $S_{avg}$ exceeds $1$ only slightly, meaning that there
are very few dimers present. Inter-particle correlations seem to be weak and
we already noted that the simulation results agree well with the theoretical
models in these cases.

As $\lambda$ is increased, the larger values of $S_{avg}$ indicate the
formation of more clusters. To elucidate this, we show snapshots of the
three-dimensional configurations in Fig.5 for the case of $\phi=0.052$
($\rho^{\ast}=0.1$) at different $\lambda$ values.  The increase of the number
and size of clusters with $\lambda$ can be seen clearly in these figures. For
$\lambda=3$ we find $S_{avg}=1.123$ at $\phi=0.052$ . About $10\%$ of the
particles are organized in dimers and less than $1\%$ in trimers. At the same
density, $S_{avg}$ increases to $1.481$ for $\lambda=4$. Correspondingly,
$22\%$ of the particles are in dimers, $7\%$ in trimers and $2\%$ in
quadrumers. The density dependence of these numbers is shown in Fig.6 where
the average percentage of particles in $n$-mers are plotted as a function of
$\phi$ for $\lambda=3$ and $4$. It can be seen that more than $60\%$ of the
particles are still in monomers and the clusters with $s \geq 5$ are rarely
found in these cases.  However, when $\lambda$ equals to $8$, all the
particles get involved in the formation of long chain-like structures. The
average chain length is larger than $13 \sigma$ at $\phi=0.052$.  Due to
thermal fluctuations, these chains are quite flexible.  They are continuously
breaking and recombining with others \cite{weis93a,levesque94a}. Even closed
ring structures can be observed occasionally. In this strong coupling case,
the chains have a close analogy to living polymers
\cite{levesque94a,tavares99a}.

When the external field is switched on, the tendency of the dipole moments to
co-align with the field direction enhances the aggregation probability.
Accordingly the average cluster size is expected to increase with the field
strength. Fig.7 shows the field-dependence of $S_{avg}$ for the systems
studied in Fig.3(a).  The increment of $S_{avg}$ with $\alpha$ is only very
weak for $\lambda\leq 2$, but it is more evident at $\lambda=4$.  As exhibited
in Fig.8, the structure character of the system varies from randomly
distributed short clusters to string-like alignments along the magnetic flux
with the increase of $\alpha$.  Consequently the cluster analysis in Fig.9
reflects that more and more particles merge into larger aggregates. Cluster as
long as $s=15$ could be observed at high $\alpha$. But these string-like
structures in Fig.8(d) are much more unstable than the chains formed at
$\lambda=8$ [see Fig.5(d)]. They posses a smaller contact energy so that
thermal fluctuations can easily break long clusters into smaller segments.
This effect is field-independent.  In fact, most of the strings observed in
Fig.8(d) consist of a few continuously breaking and recombining segments. As a
result, the average cluster size $S_{avg}$ does not present a dramatic change
as expected from the snapshot of the configurations.  When $\lambda=8$,
however, the long-chains had already been formed even at zero field. The
increase of the field strength just arrange these chains to have a better
alignment along the field direction. But the average chain length does not
change too much.

The results in Fig.7 reveal that the magnetic field dependence of
$S_{avg}$ remains weak at small $\alpha$.  It is therefore legal
to invoke the zero-field cluster analysis in order to interpret
the outcomes related to the initial susceptibility (Fig.2). Let
us focus to the cases of $\lambda=3$ and $4$ at low
concentrations where the deviation between theory and simulation
are most pronounced. In these cases the small clusters are
relatively stable and spatially well separated from each other.
For a rough approximation, we treat the clusters as rigid
particles thus considering our system as a poly-dispersed sample.
We define the effective dipole moment of each cluster according to
\begin{equation}
\VECm_s=\sum_{i=1}^{s}\VECm_i.
\end{equation}
The effective Langevin susceptibility $\chi_{L}^{eff}$ of such a
poly-dispersed system can be computed as
\begin{equation} \label{Langsuseff}
\chi_L^{eff}=\frac{1}{V}\frac{<\sum_{all\ clusters} m_s^2>}{3\mu_0 kT}.
\end{equation}
Since most of the clusters are short linear $n$-mers, the value
of $\chi_{L}^{eff}$ is larger than $\chi_L$ - the appropriate
value for the mono-dispersed system. Replacing $\chi_L$ by
$\chi_{L}^{eff}$ in Eq.(\ref{sus3ord}) we get new modified
theoretical curves. In Fig.10 it can be seen that this simple
operating prescription considerably improves the agreement
between simulation and theory, at least at low concentrations.
This proves that the formation of small clusters is responsible
for the enhancement of the initial susceptibility.

At particle concentration above $\phi \simeq 0.1$, the modified theoretical
curves start to overestimate the simulated $\chi$-values. In these cases, the
increased angular correlations between the particles are not just limited to
the two next nearest neighbors and thus make the above definition of clusters
ambiguous \cite{levesque94a}. Although we keep on using the energy criterion
to obtain the cluster informations for Fig.4 and 6, these values are more
likely reflecting the local orientational order of the dipole moments. The
clusters we find are not stable entities involving well-defined particle
numbers as in the low density regime. This can be partially seen from the
$S_{avg}(\phi)$ behavior in Fig.4. When $\lambda\leq 2$, $S_{avg}$ slightly
increases with $\phi$. But it saturates at $\lambda=4$ and even decreases at
$\lambda=8$ for $\phi \geq 0.1$.  Next we examine the radial distribution
function $g(r)$ of the system (Fig.11), which gives the probability of finding
a pair of particles a distance $r$ apart, relative to the probability expected
for a completely random distribution at the same density\cite{allen87a}. We
find that the height of the first and second peaks, respectively, associated
with dimer and trimers, decrease with the increase of $\phi$. This implies
that the number of clusters does not increase as rapidly as the density
\cite{stevens95a}. In other words, upon rising $\phi$ the spatial distribution
of the particle positions re-approaches to that of a homogeneous system.
Therefore we expect our suggested modification with Eq.(\ref{Langsuseff}) only
to be valid at low volume fractions below $\phi = 0.1$.

The formation of long chain-like structures in the strong coupling regime
makes the systematic investigation of the magnetization properties at
$\lambda=8$ difficult.  At $\phi = 0.0196$, there are very few overlaps
between the elongated chains.  They have enough free volume to re-arrange in
response to the external field. The magnetization of the system increases very
fast in the weak field regime, as shown in Fig.3(a). The occasional formation
of a few closed ring structures is not found to affect noticeably the global
magnetization behavior. On the other hand, when $\phi$ is enlarged, the chains
become strongly entangled, like in a semi-flexible polymer solution. This
slows down the dynamics of the system dramatically. An extremely long
simulation time is required to reach the equilibrium state in the weak field
regime. For example, at a density of $\phi=0.0785$, the equilibrium value of
$\VECM$ at $\alpha=0.5$ is obtained only after 1.5 million simulation time
steps.  Simultaneously, since the number of particles is fixed in our
simulation, the increase of $\phi$ makes the chain length become comparable to
the size of the simulation box. This unavoidably induces finite size effects.
For these reasons, we were unable to provide systematic investigations at
$\lambda=8$ at higher concentrations (or equivalently larger $\chi_L$), but
only for $\chi_L = 1.256$ shown in Fig.3(a).  These numerical difficulties
could well be related to experimental observations which find a dramatic
increase of viscosities at similar dipolar couplings \cite{odenbachpc}.

\subsection{Comparison with ferro-solids}

To further investigate the influence of the cluster formation process we now
turn to ferro-solid systems. By this we denote a suspension of dipolar
particles which have their spatial positions frozen in, but are permitted to
rotate freely. In this way the aggregation related effects in ferrofluids can
easily be separated.

We prepared the initial configurations of the ferro-solid system by placing
the particles randomly in the simulation box with random dipole moment
orientations. For each given $\lambda$ and $\phi$, several different initial
configurations were generated and simulated. They were found to give
very similar results. Computer time limitations prohibited us to
average over a large
sequence of quenched configurations, which would be desirable to obtain a good
ensemble average for a ferro-glass system. However, we have investigated this
case only to study the qualitative influence of freezing the positional
degrees of freedom. Accordingly, only the rotational equations of motion
[Eq.(\ref{LangevinR})] of the particles need to be integrated in the
simulations. Fig.12 displays the simulation results for the initial
susceptibility at $\lambda=3$ and $4$, along with the theoretical predictions
for the fluid system. In ferro-solids $\chi$ also exhibits a nonlinear
increase with the volume fraction $\phi$. Comparing our data with the results
of the ferrofluid system in Fig.2, we see that the suppression of clusters
leads to a slower increase of the magnetization at weak fields, and hence to a
lower $\chi$. It is interesting to note that the discrepancy between
$\chi_{solid}$ and $\chi_{fluid}$ goes up when $\lambda$ or $\phi$ are
increased. This shows that the positional correlations are also very important
at high density, and couple strongly to the dipolar rotational degrees of
freedom, even if they do not evidently form stable clusters. We consider the
fact that the measured values of $\chi_{solid}$ coincide with the theoretical
curves for the fluid system, Eqs.(\ref{sus2ord}) and (\ref{sus3ord}), almost
exactly at low concentration for an accidental correspondence. At higher
volume fractions the ferro-solid displays a susceptibility even below the
theoretical prediction.  This effect is more evident for larger $\lambda$ at
higher $\phi$. Note, that in Ref.  \onlinecite{pshenichnikov00a} a similar
solidified ferro-colloid was investigated. There $\chi_{solid}$ was found to
increase with $\chi_L$, but the values are larger than that for the associated
ferrofluid. Again we suspect that this due to finite size effects of their
sample.

We point out that the particles in our ferro-solid systems are supposed to be
super-paramagnetic. This means that they have a negligible crystallographic
magnetic anisotropy energy. There is no energy barrier which prevents the
magnetic dipoles from a free rotation. That makes the present systems
different from that one studied in Ref.\onlinecite{chantrell00a} where the
initial susceptibility is found to decrease with the increase of the particle
concentrations.

\section{Conclusion}
We have investigated in detail the initial susceptibility, magnetization curve
and microstructure of ferrofluids in various particle concentration and
dipolar coupling strength ranges by means of molecular dynamics simulations.
The ferrofluid was modeled as a soft sphere system by means of a purely
repulsive Lennard-Jones potential, and the magnetic cores interacting with a
dipolar coupling strength $\lambda$ from $1$ to $8$.  The investigated volume
fractions were in the range of $\phi \leq 0.24$ mostly. Our simulation method
used the Ewald summation with metallic boundary conditions to deal with the
long-range dipolar interactions, and took explicitly into account the
translational and rotational degrees of freedom of the dipolar particles. The
temperature was kept constant by means of a Langevin thermostat for all
degrees of freedom. Our simulation results for the susceptibility show good
agreement with the theoretical predictions in the weak coupling limit $\lambda
\le 2$. For higher dipolar couplings, however, we find systematic deviations
from the theoretical predictions at low and intermediate concentrations. The
detailed cluster analysis demonstrates that this is due to the particle
aggregation at higher $\lambda$.  The formation of clusters tend to increase
the magnetization at weak field regime and consequently induce a larger
initial susceptibility.  At high densities, the spatial distribution of the
particles starts to homogenize again, and the significance of clusters goes
down.  In fact we speculate that at higher concentrations there are almost no
stable clusters any more, because they can disintegrate and reform with other
particles rapidly. In this regime we find that the third order equation in
$\chi_L$ is compatible with our data.

For the magnetization curves we find that both, the second and
third order theoretical curves, describe our data well for
couplings $\lambda \le 2$. For higher values of $\lambda$ we find
again systematic deviations.  The system shows at coupling
strength of $\lambda = 8$ an almost gel-like behavior due to
entangled chain structures. We expect this effect to become
experimentally relevant in Cobalt based ferrofluids, where
$\lambda$ is large.

The influence of the aggregation phenomena was isolated by
studying ferro-solids where the particles are randomly placed
inside the simulation volume, and only rotations are allowed to
occur. The observed initial susceptibilities are lower compared
to the ferrofluid systems due to the suppression of cluster
formation. The deviations to the fluid system grow with
increasing dipolar coupling strength and increasing volume
fraction.

\section*{Acknowledgments}

We thank B. Huke and S. Odenbach for helpful discussions. Financial support
from the DFG under grant No. HO 1108/8-2 is greatly appreciated.

\newpage
\begin{table}
\caption{Linear regression fitting results for the M(H) [M($\alpha$)] curves
  in Fig.1. The theoretical predictions of $\chi$ on different orders
  are also given for comparison.}
\vspace{1cm}
\begin{tabular}{|c|c|c|c|c|c|c|c|c|}
\hline
$\phi$ & $\chi$ & $\Delta \chi$ & $\xi^2/(N_{d,opt}-2)$ & $Q$ &
$N_{d,opt}$&$\chi_L[Eq.(\ref{Langsus})]$&$\chi[Eq.(\ref{sus2ord})]$ & $\chi[Eq.(\ref{sus3ord})]$\\
\hline
0.0262 & 0.822 & 0.017 & 0.069 & 0.999 & 11 & 0.6283 & 0.7599 & 0.7616\\
\hline
0.0524 & 2.040 & 0.045 & 0.138 & 0.997 & 10 & 1.2566 & 1.7830 & 1.7968\\
\hline
0.0785 & 3.505 & 0.069 & 0.249 & 0.981 & 10 & 1.8849 & 3.0693 & 3.1158\\
\hline
0.1047 & 5.103 & 0.142 & 0.085 & 0.999 & 9 & 2.5133 & 4.6188 &  4.7290\\
\hline
0.1309 & 7.173 & 0.330 & 0.069 & 0.997 & 7 & 3.1416 & 6.4315 &  6.6468\\
\hline
0.1571 & 9.237 & 0.363 & 0.094 & 0.993 & 7 & 3.7699 & 8.5073 &  8.8794\\
\hline
0.1833 & 11.680 & 0.741 & 0.016 & 0.999 & 6 & 4.3982 & 10.8464 & 11.4372\\
\hline
0.2094 & 14.356 & 0.873 & 0.140 & 0.968 & 6 & 5.0266 & 13.4486 & 14.3306\\
\hline
0.2356 & 17.831 & 1.114 & 0.150 & 0.963 & 6 & 5.6549 & 16.3140 & 17.5698\\
\hline
\end{tabular}
\end{table}
\clearpage
\newpage
\vspace{3cm}
\begin{figure}[htb]
\begin{center}
\includegraphics*[height=10.0 cm]{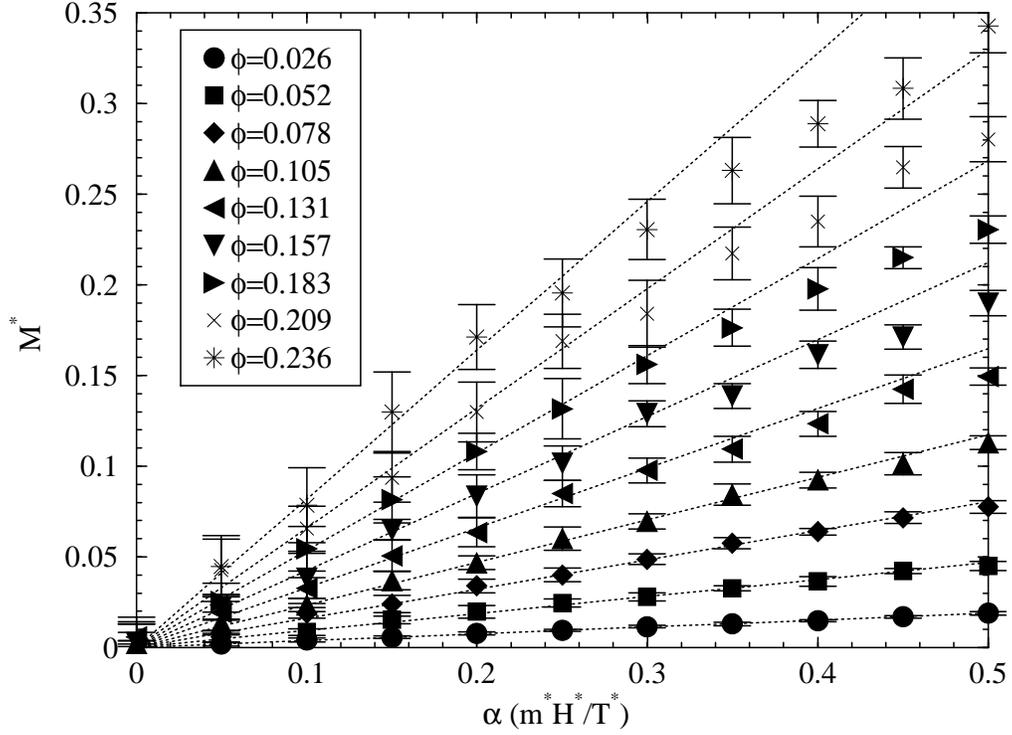}
\end{center}
\caption{Dimensionless magnetization $M^{\ast}$ as a function of
the Langevin parameter $\alpha$ and the volume fraction $\phi$
for a system with $\lambda=3$. The symbols are the simulation
results. The dashed curves are the linear fitting data, the slope
of which  gives the initial susceptibility $\chi$.}
\label{fig1} 
\end{figure}
\newpage
\vspace{3cm}
\begin{figure}[htb]
\begin{center}
\includegraphics*[height=9.0cm]{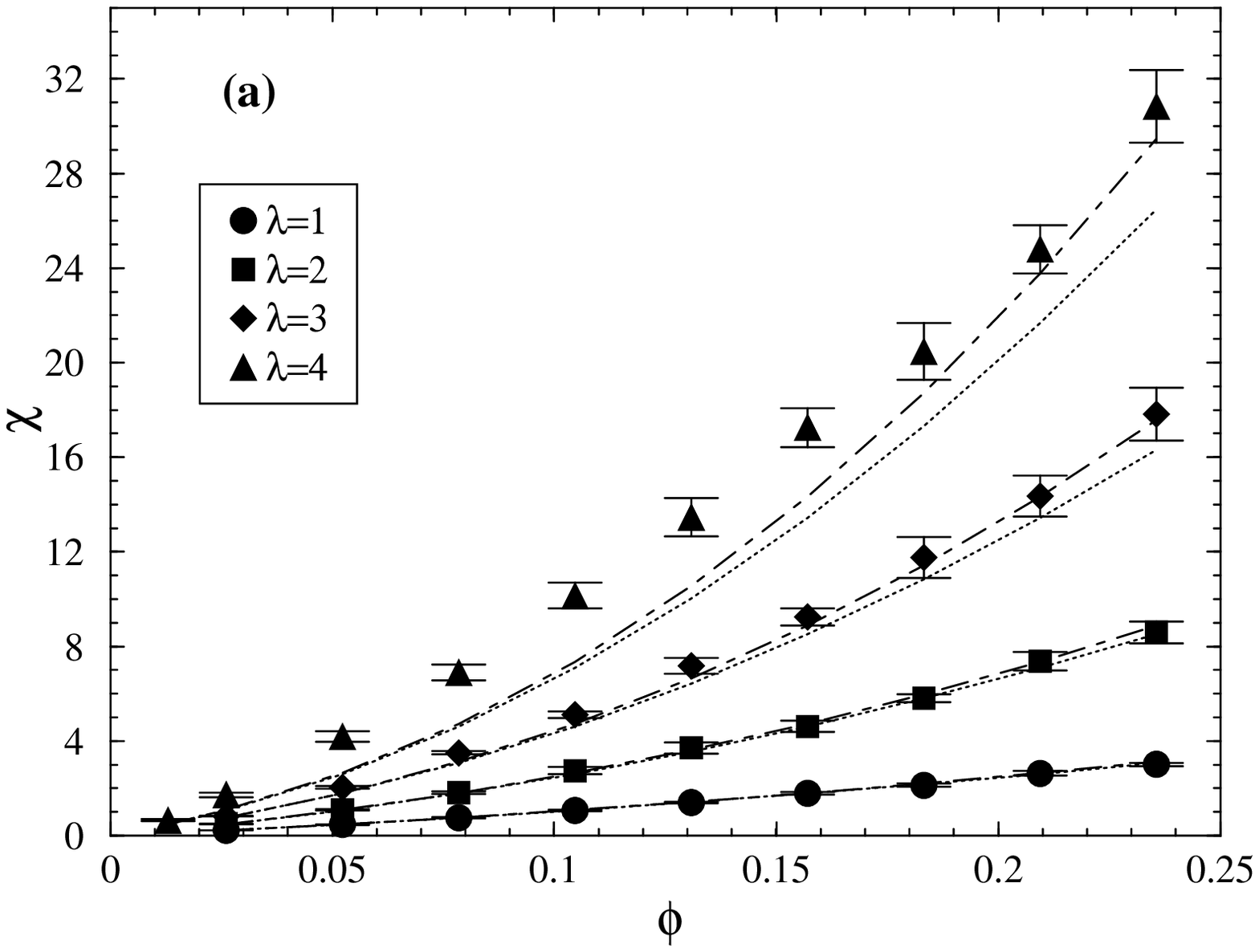}
\includegraphics*[height=9.0cm]{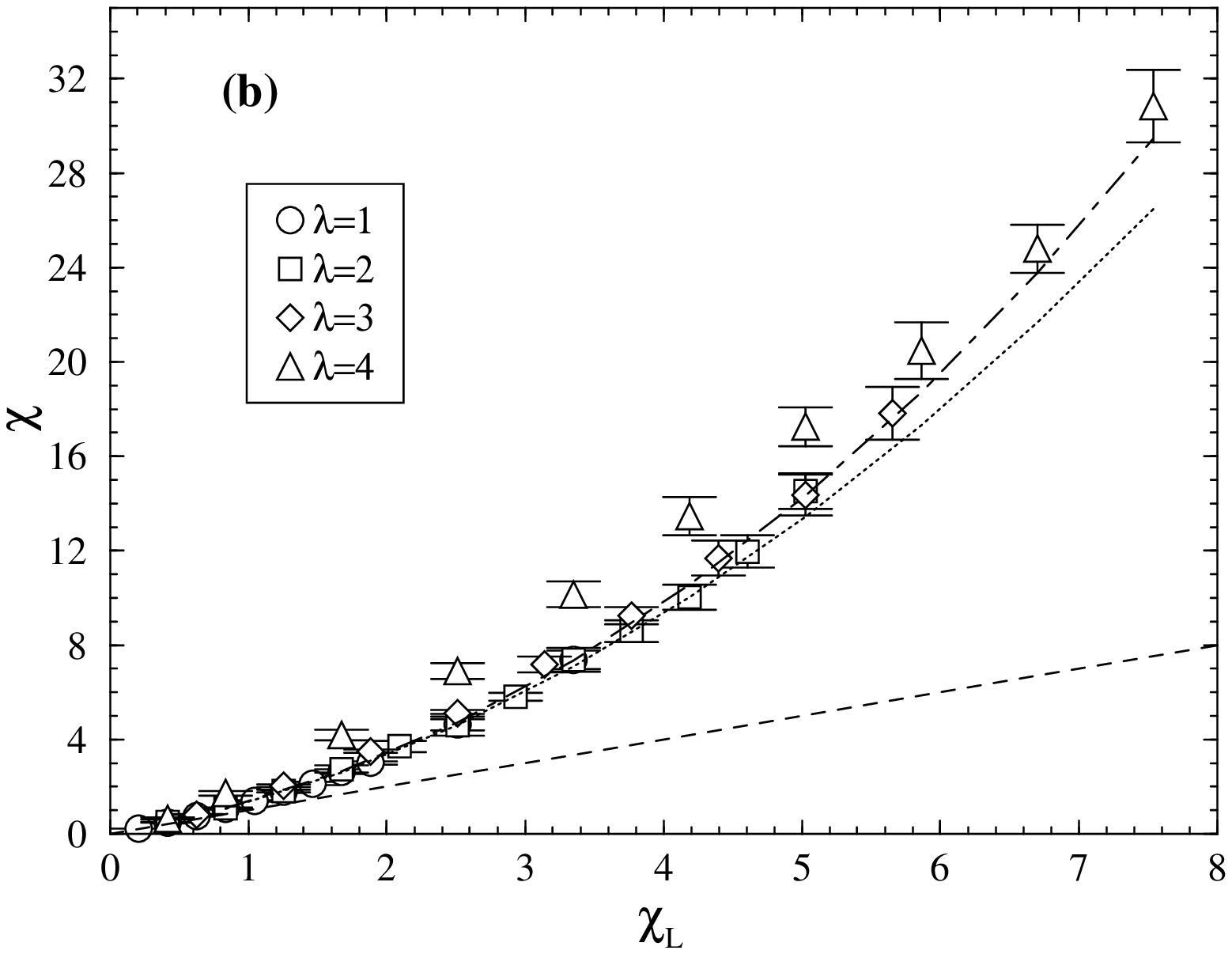}
\end{center}
\caption{(a) Initial susceptibility $\chi$ as a function of the
dipolar coupling constant $\lambda$ and volume fraction $\phi$;
(b) same data of $\chi$ as in (a), but mapped on the Langevin
susceptibility $\chi_L$. The theoretical curves are given by
the Langevin model ($\chi=\chi_L$, dashed), Eq.(\ref{sus2ord}) (dotted) and
Eq.(\ref{sus3ord}) (dotted-dashed), respectively.}
\label{fig2} 
\end{figure}
\newpage
\begin{figure}[htb]
\begin{center}
  \makebox[0cm][t]{\hspace{5.2cm}\parbox[t]{0cm}
    {\vspace{-7.cm}\includegraphics*[height=5.8cm]{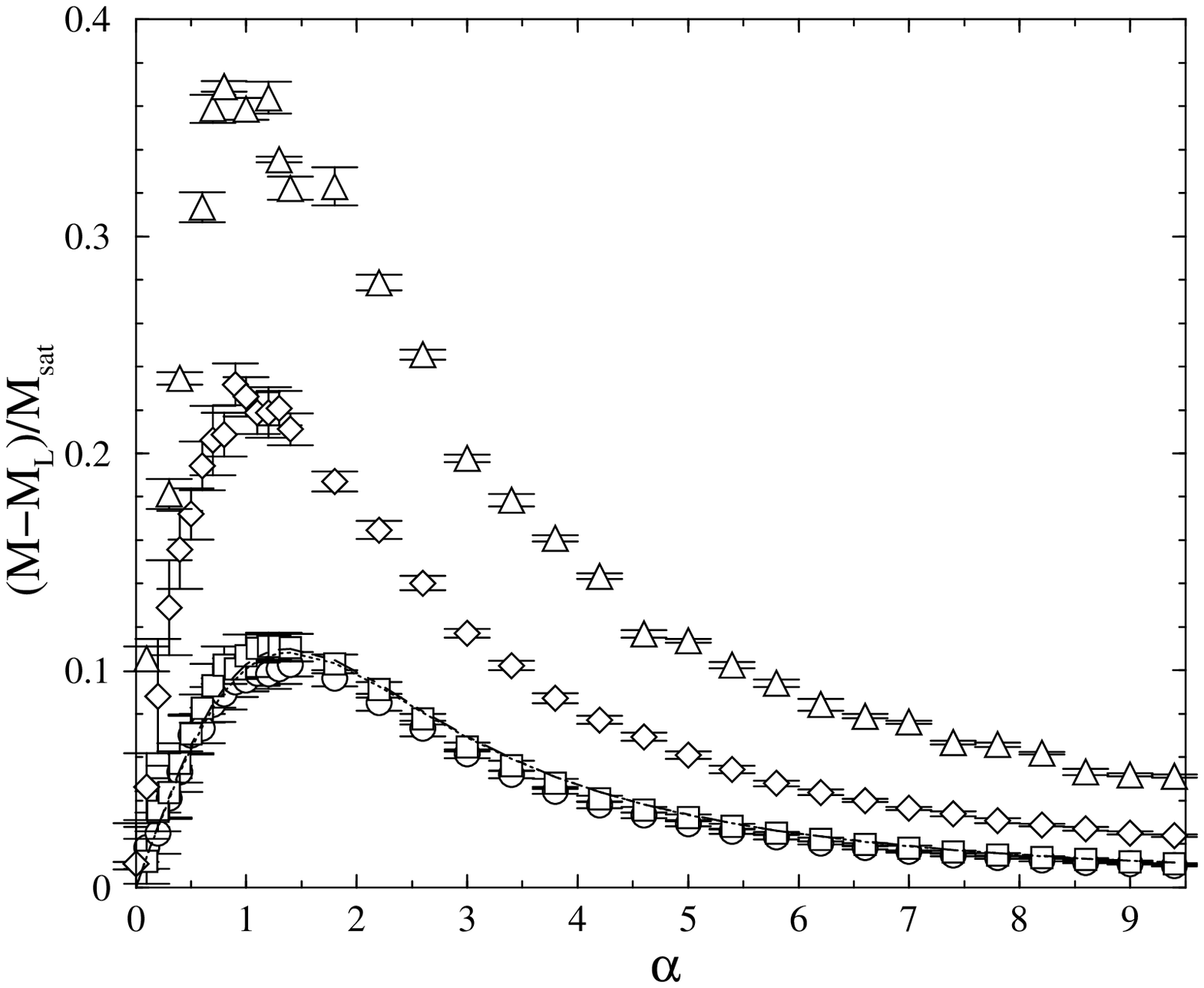}}}
  \includegraphics*[height=9.5cm]{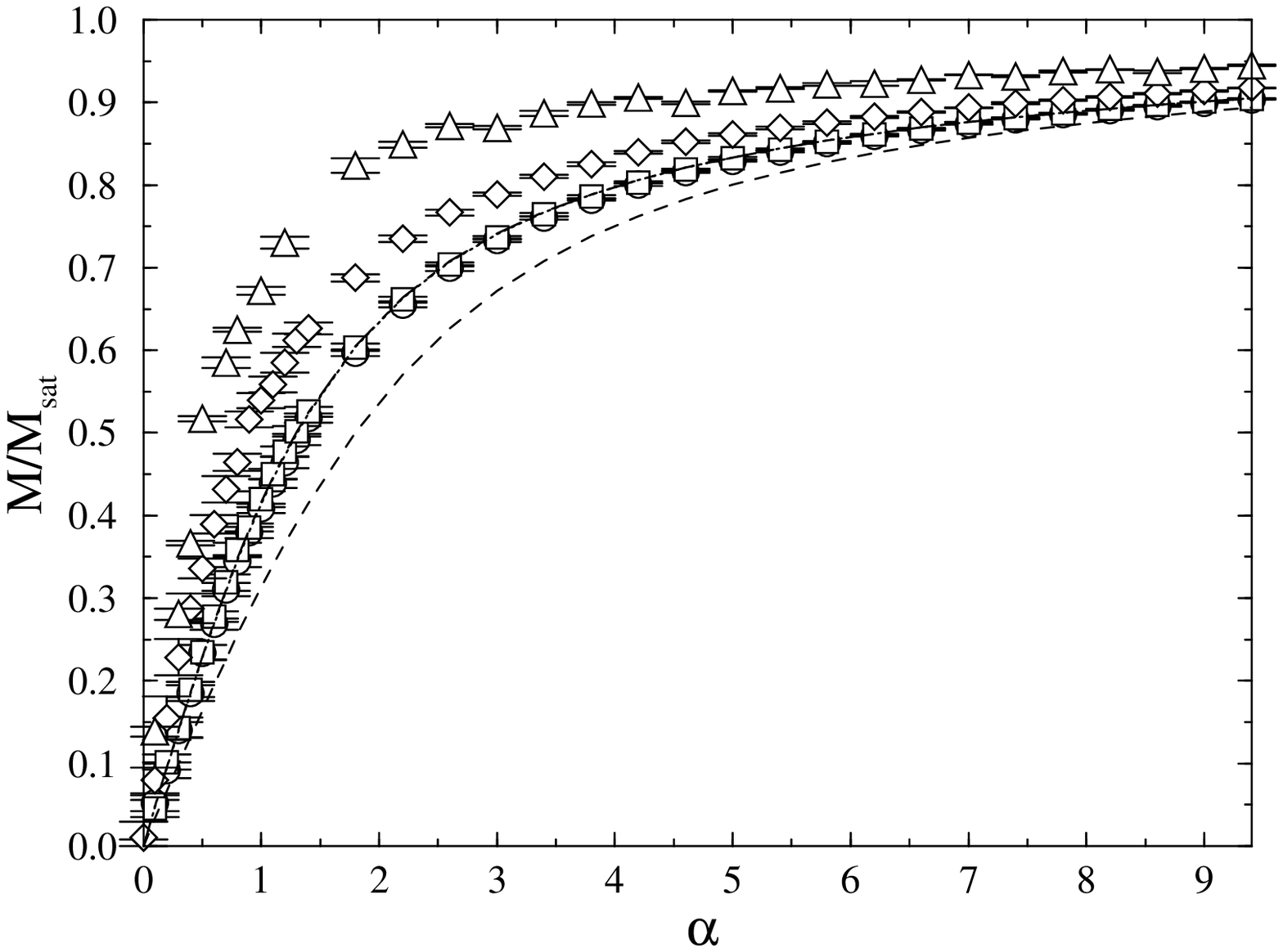}

  \makebox[0cm][t]{\hspace{5.2cm}\parbox[t]{0cm}
    {\vspace{-7.cm}\includegraphics*[height=5.8cm]{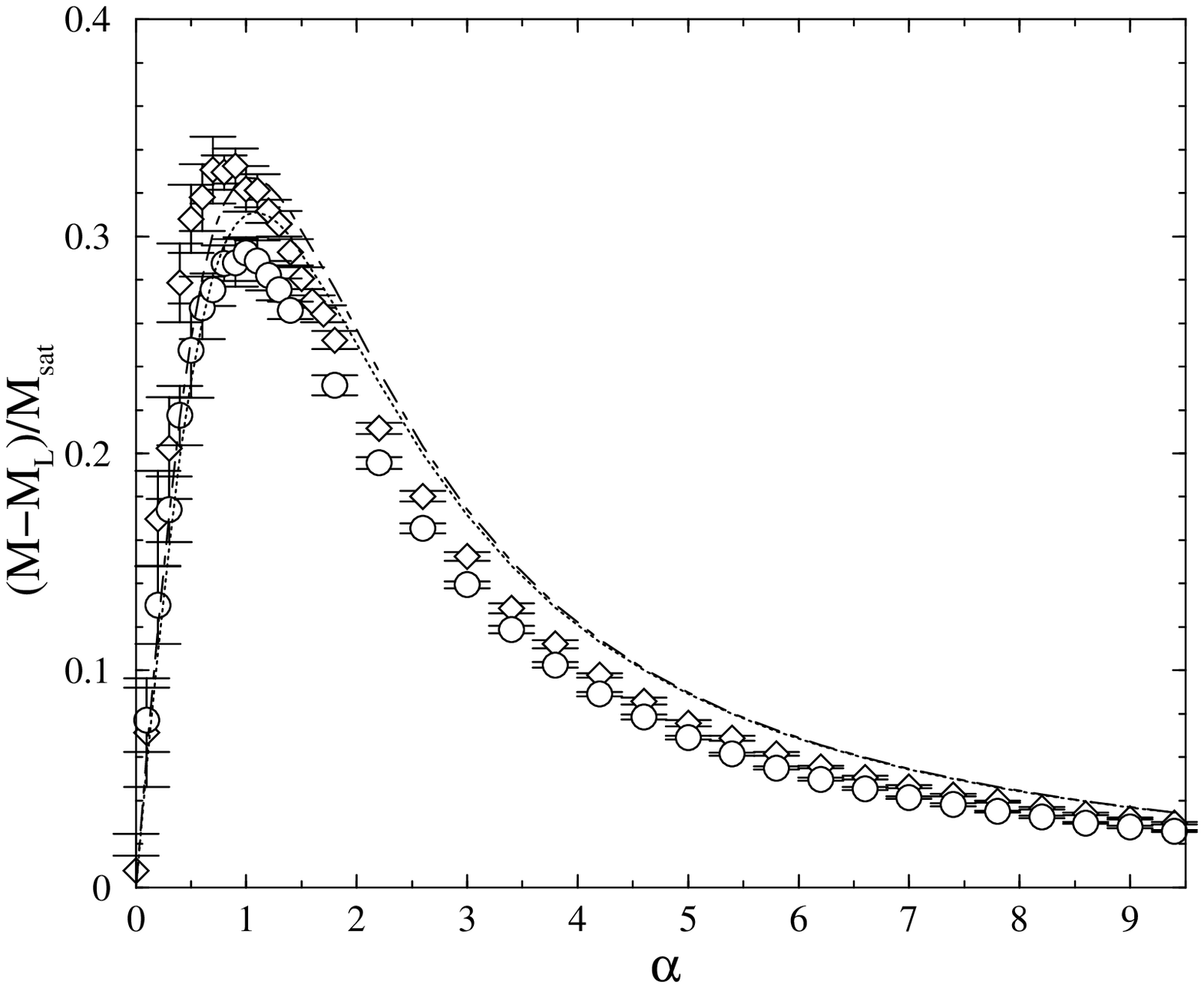}}}
  \includegraphics*[height=9.5cm]{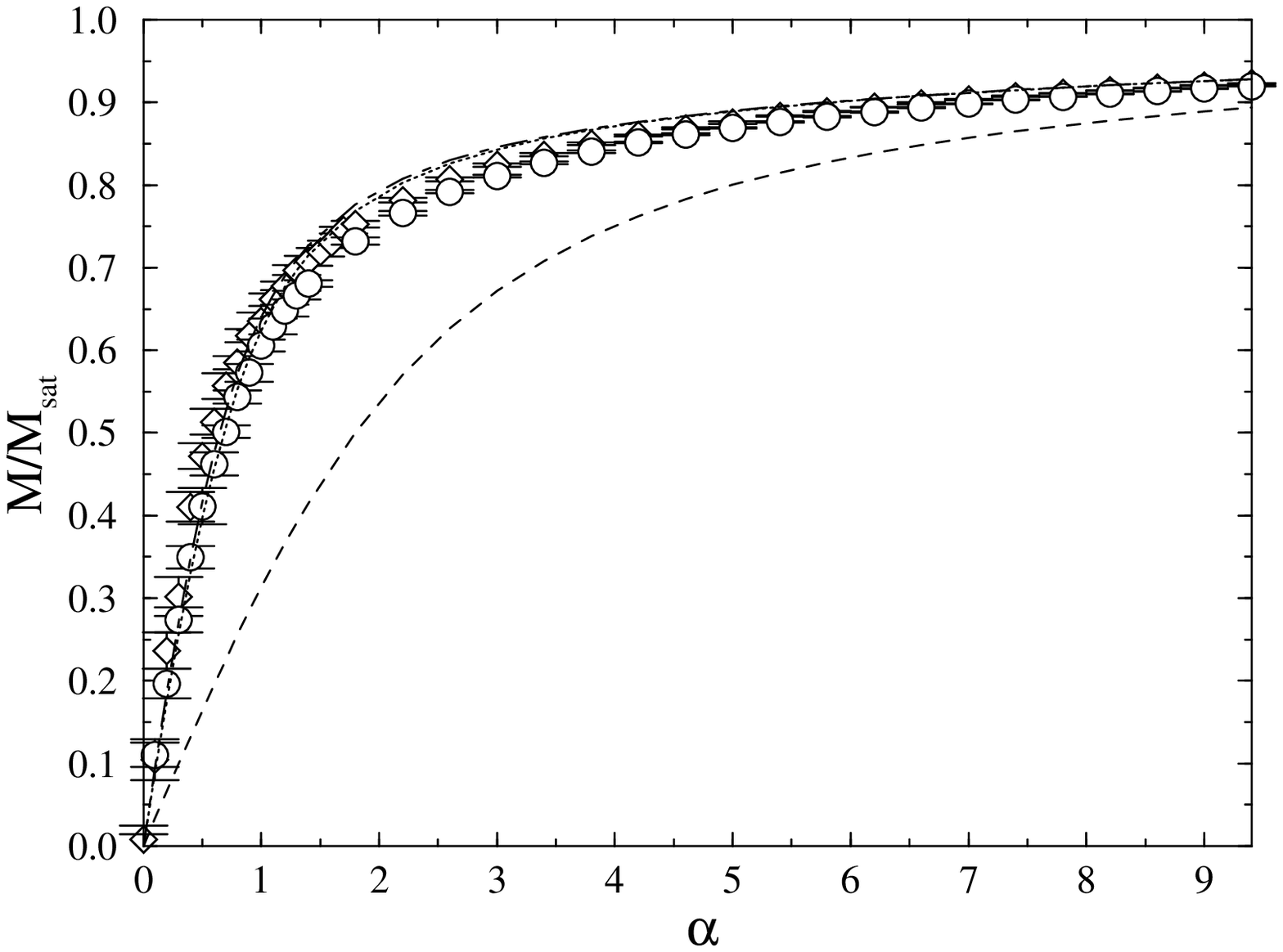}
\end{center}
\caption{Magnetization curves in case of (a) $\chi_L=1.256$
  for $\lambda=1$ (circle), $2$ (square), $4$ (diamond) and $8$ (triangle-up);
  (b) $\chi_L=5.026$ for $\lambda=3$ (circle) and $4$ (diamond), respectively.
  The theoretical curves are given by the Langevin model
  [Eq.(\ref{LangevinM}), dashed line], and by using Eq.(\ref{magz2ord})
  (dotted line) and Eq.(\ref{magz3ord}) (dotted-dashed line) in
  Eq.(\ref{magzthe}), respectively.  The inlays show the difference between
  the calculated magnetization and the Langevin model $(M-M_L)/M_{sat}$.}
\label{fig3} 
\end{figure}
\newpage
\vspace{3cm}
\begin{figure}[htb]
\begin{center}
\includegraphics*[height=10cm]{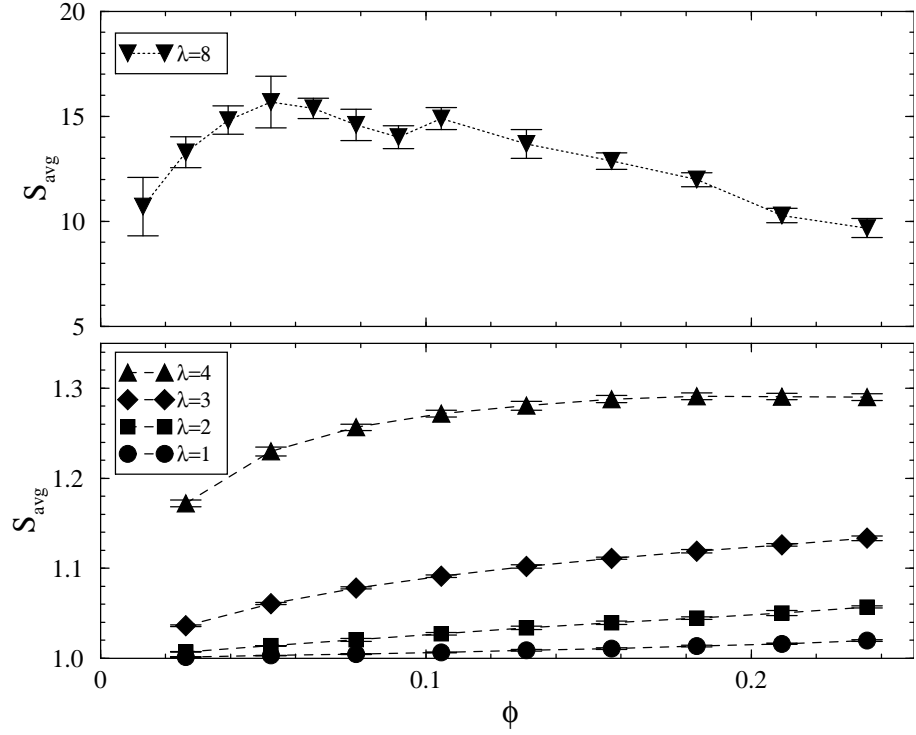}
\end{center}
\caption{Average size of the clusters formed at zero
field as a function of $\lambda$ and $\phi$.}
\label{fig4} 
\end{figure}
\newpage
\vspace{3cm}
\begin{figure}[htb]
\begin{center}
\includegraphics*[width=0.46\textwidth]{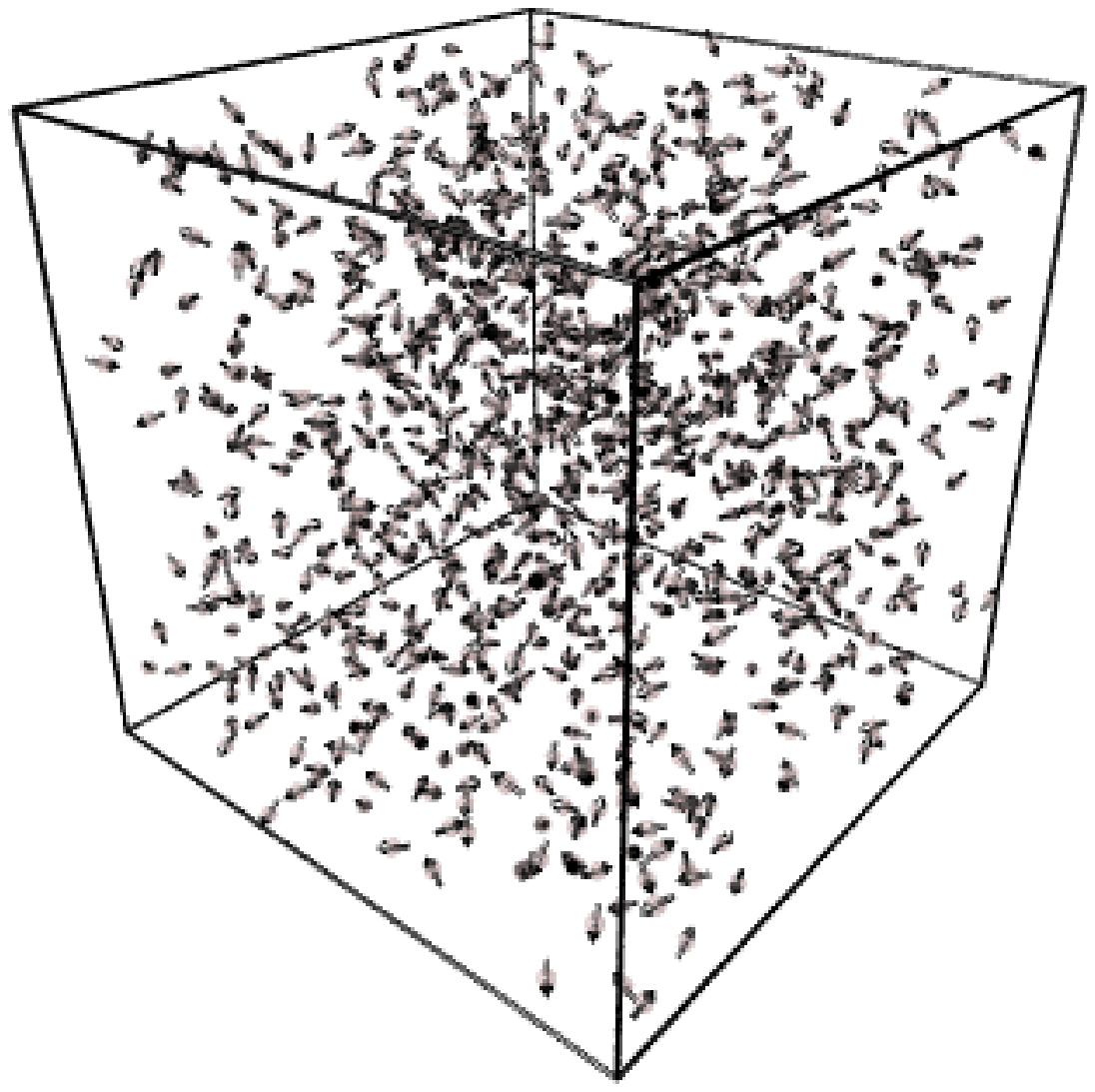}
\includegraphics*[width=0.46\textwidth]{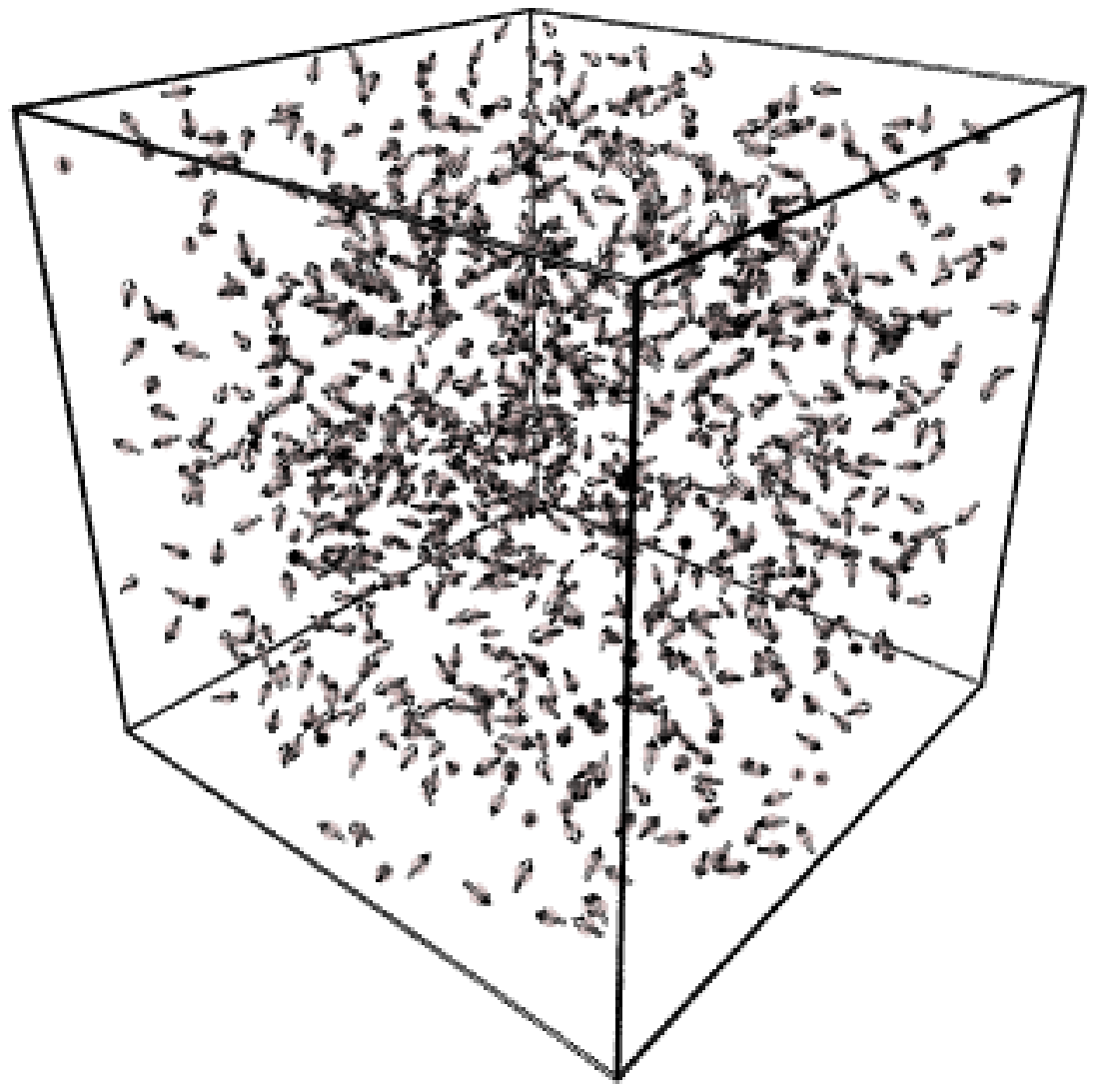}
\includegraphics*[width=0.46\textwidth]{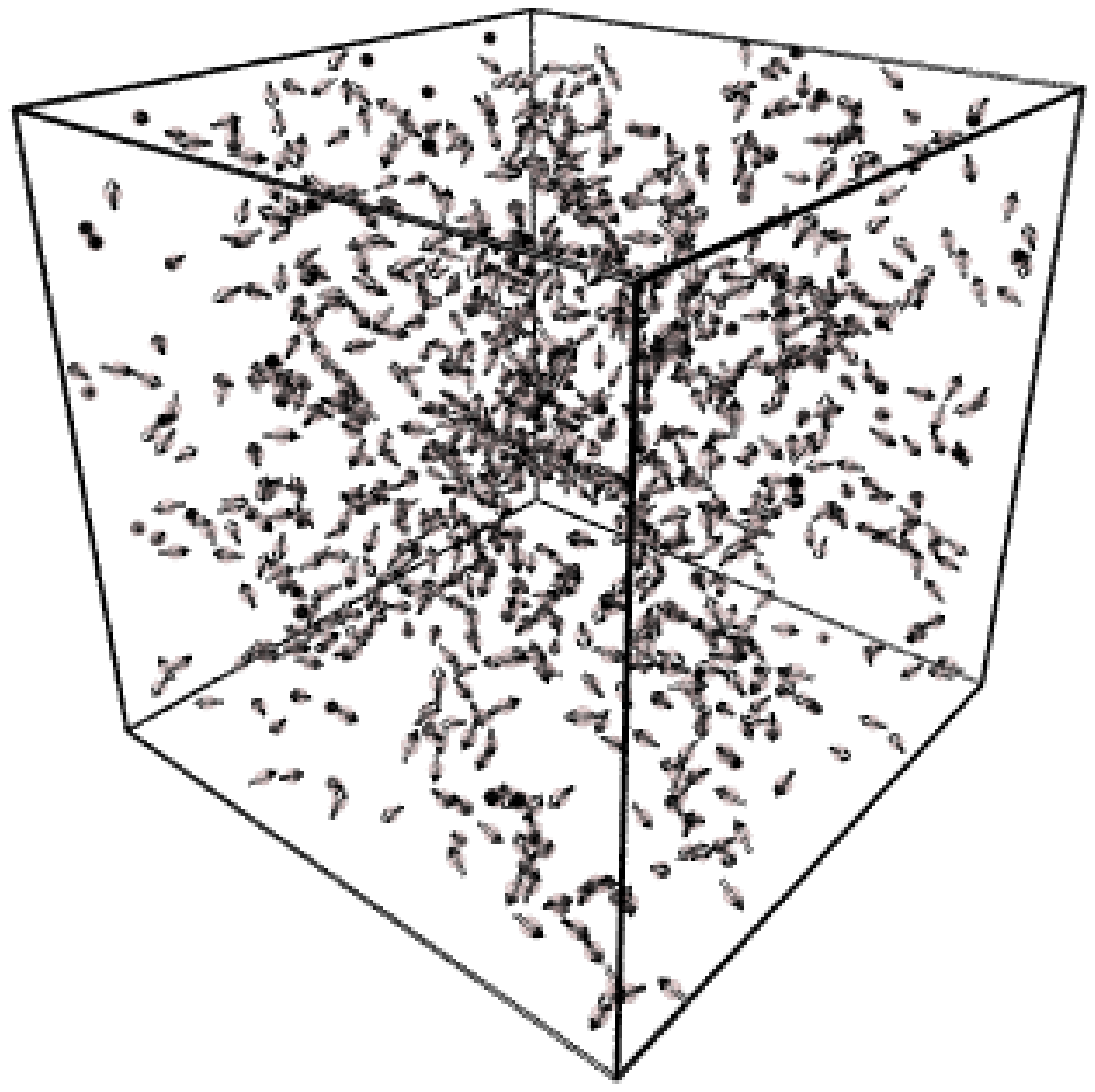}
\includegraphics*[width=0.46\textwidth]{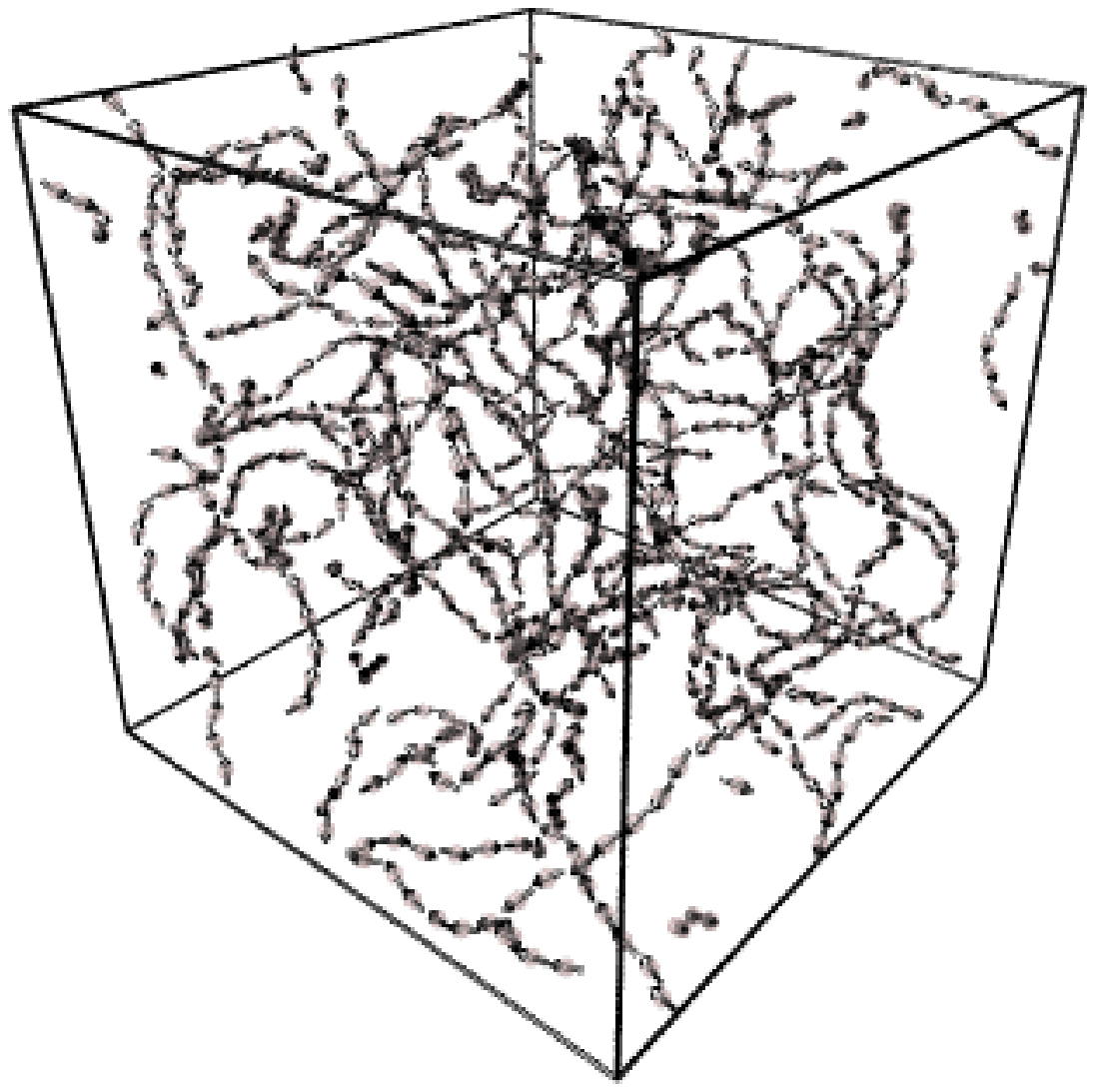}
\end{center}
\caption{Snapshot of the three-dimensional configurations formed
at zero-field in the systems of $\phi=0.052$ with $\lambda=1$
(a), $3$ (b), $4$ (c) and $8$ (d), respectively. The arrows with
length of $\sigma$ indicate the orientation of the dipole moments.
The spheres (diameter=$\sigma/2$) show the center positions of
the particles.}
\label{fig5} 
\end{figure}
\newpage
\vspace{2cm}
\begin{figure}[htb]
\begin{center}
\includegraphics*[height=9.0cm]{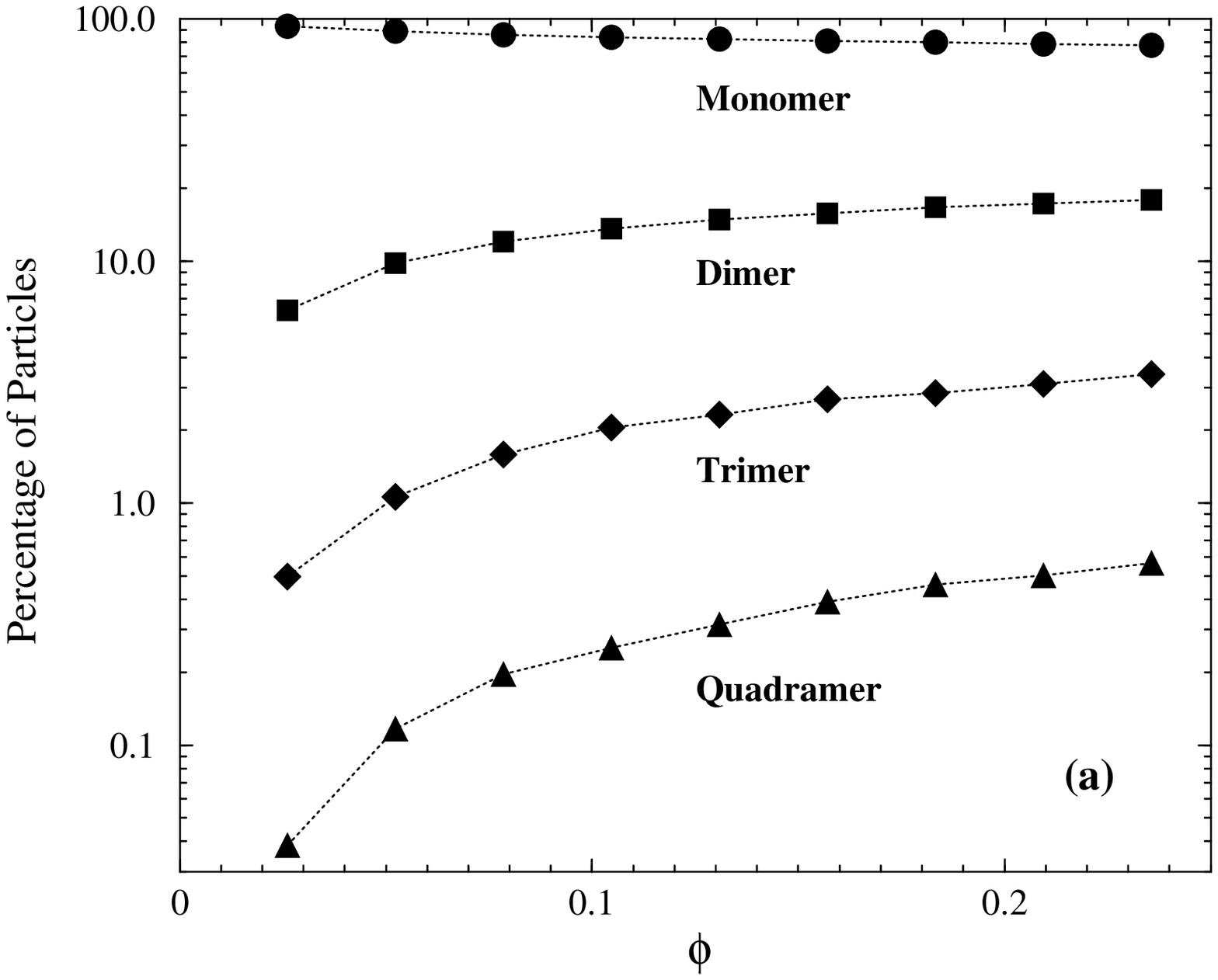}
\includegraphics*[height=9.0cm]{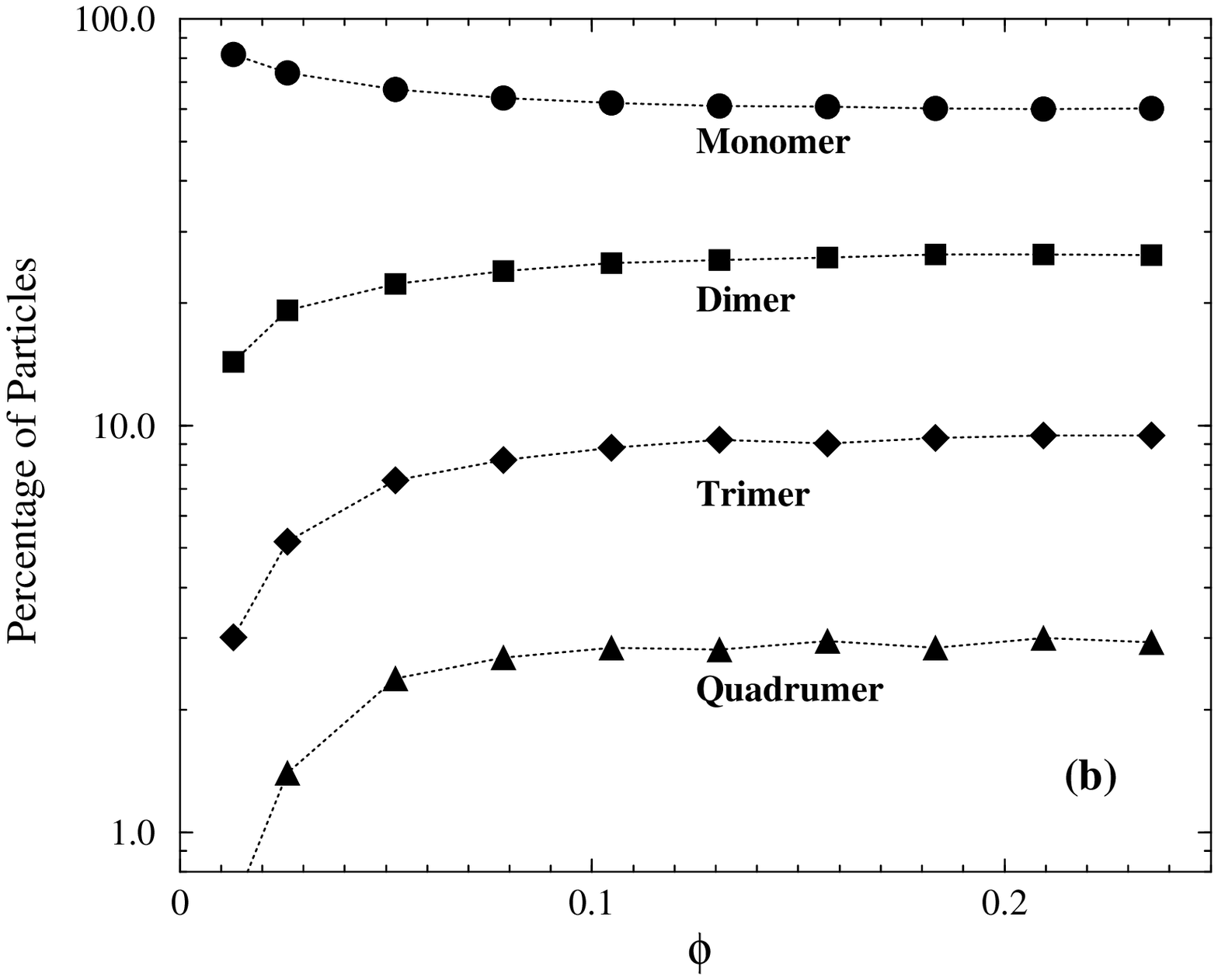}
\end{center}
\caption{Average percentage of the particles in $n$-mers
for the system with $\lambda=3$ (a) and $4$ (b) at zero-field.}
\label{fig6} 
\end{figure}
\newpage
\vspace{4cm}
\begin{figure}[htb]
\begin{center}
\includegraphics*[height=10cm]{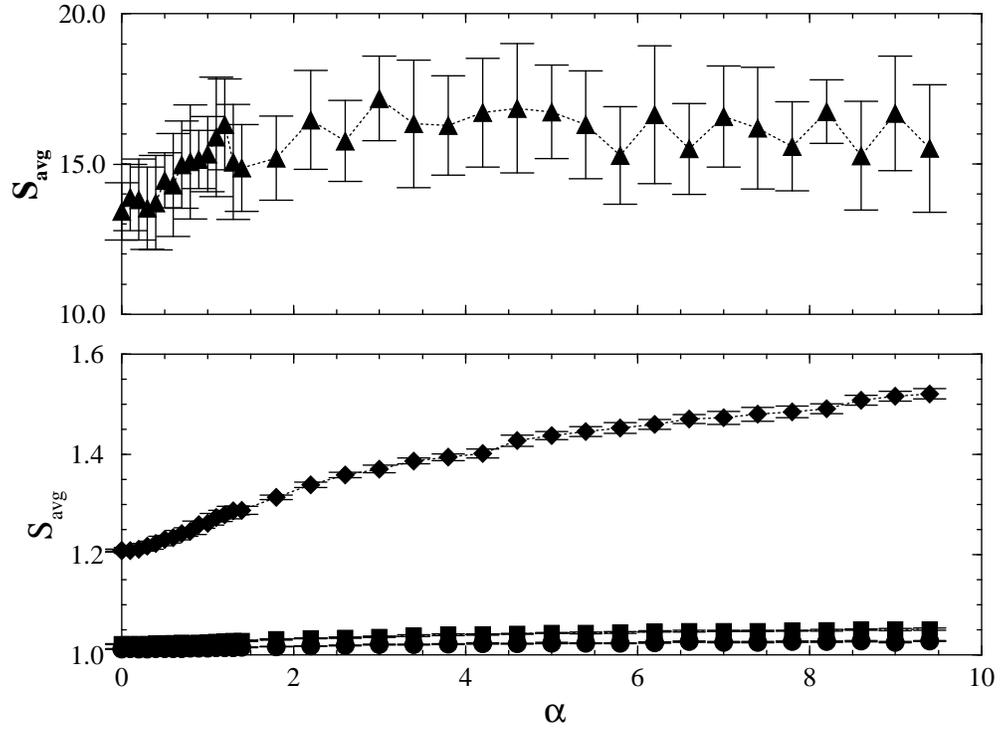}
\end{center}
\caption{Field-dependence of the average cluster size
of the systems studied in Fig.3(a): $\lambda=1$ (circle),
$2$ (square), $4$ (diamond) and $8$ (triangle-up).}
\label{fig7}
\end{figure}
\newpage
\vspace{3cm}
\begin{figure}[htb]
\begin{center}
\includegraphics*[width=0.46\textwidth]{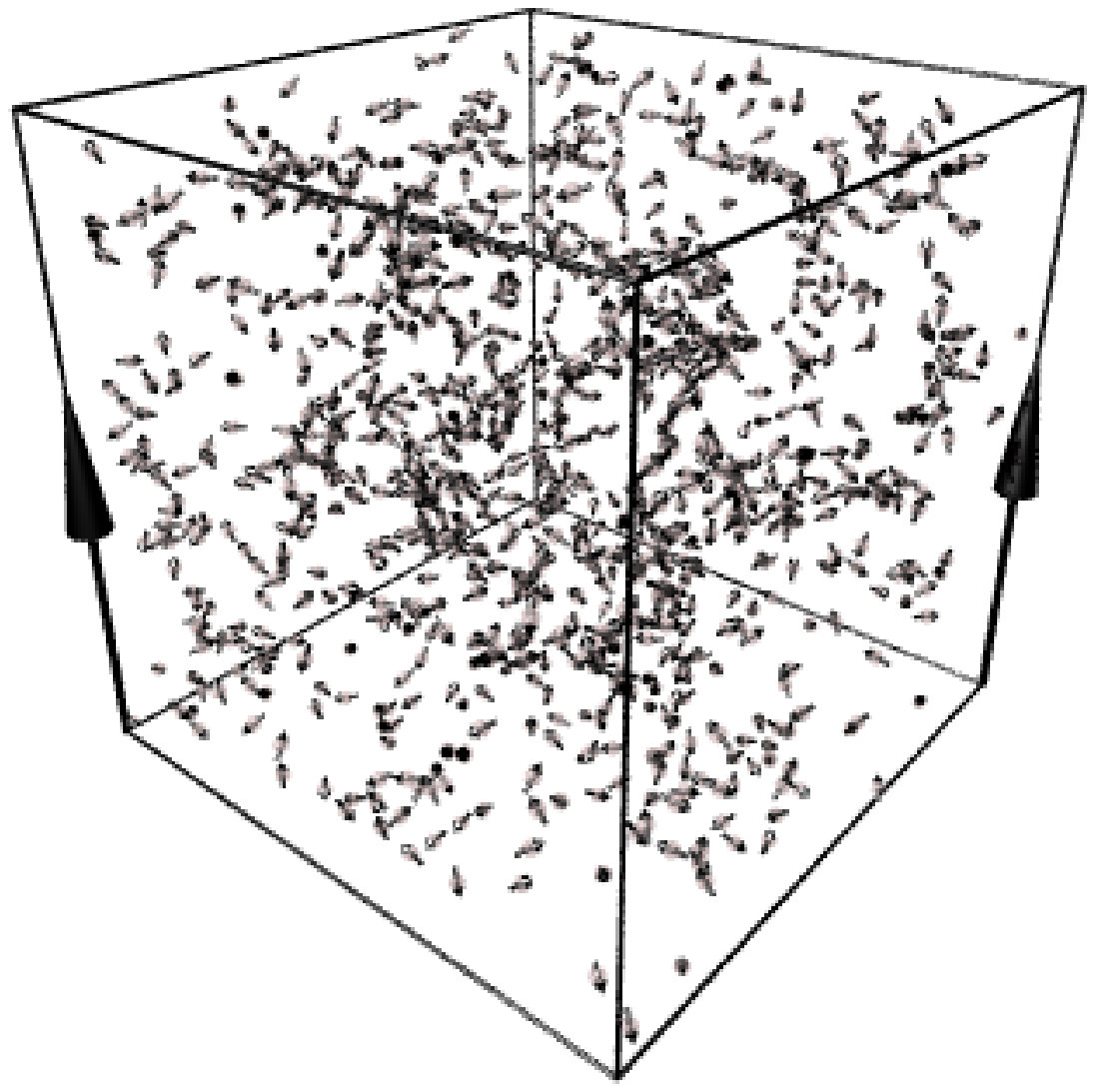}
\includegraphics*[width=0.46\textwidth]{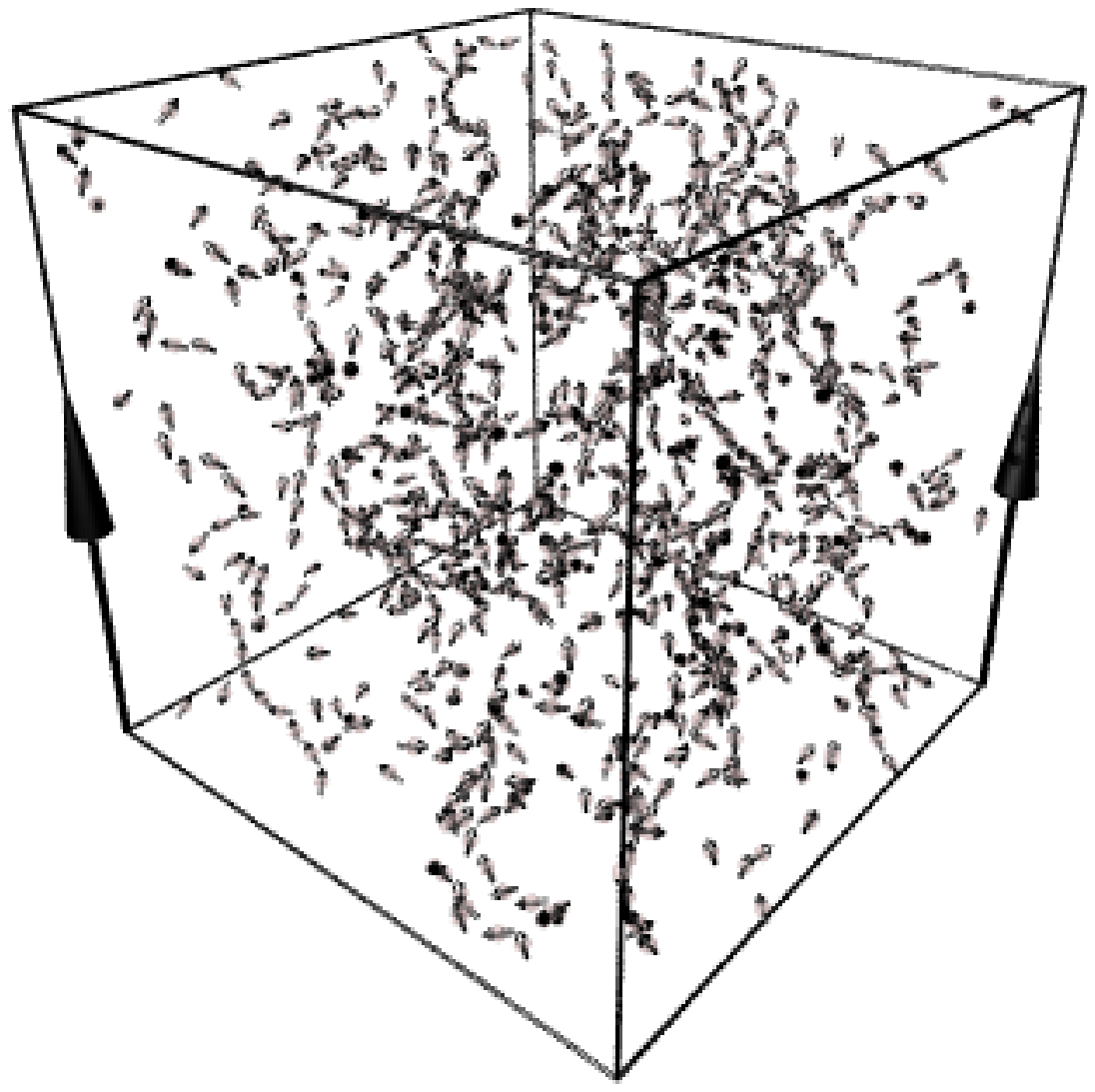}
\includegraphics*[width=0.46\textwidth]{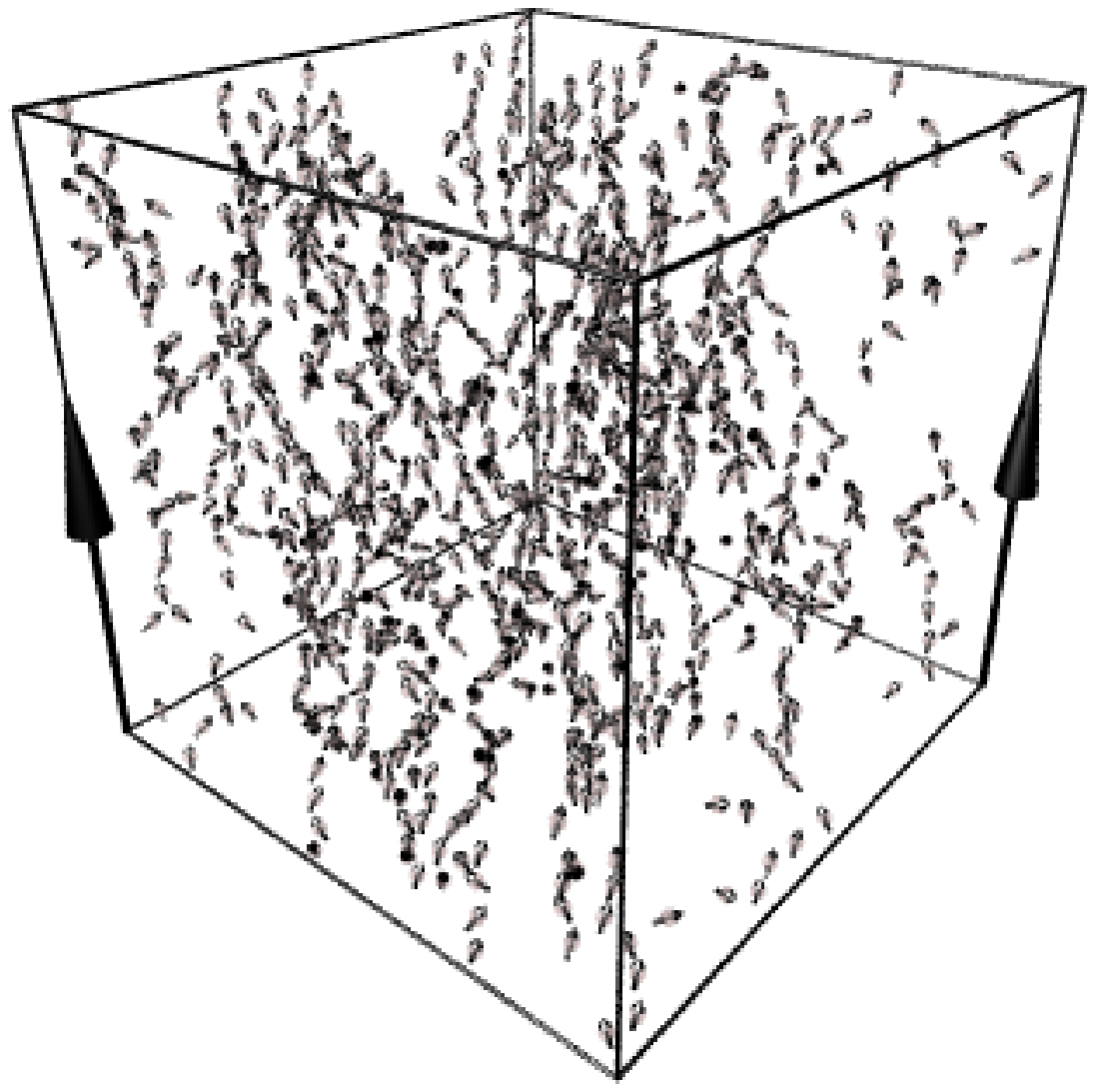}
\includegraphics*[width=0.46\textwidth]{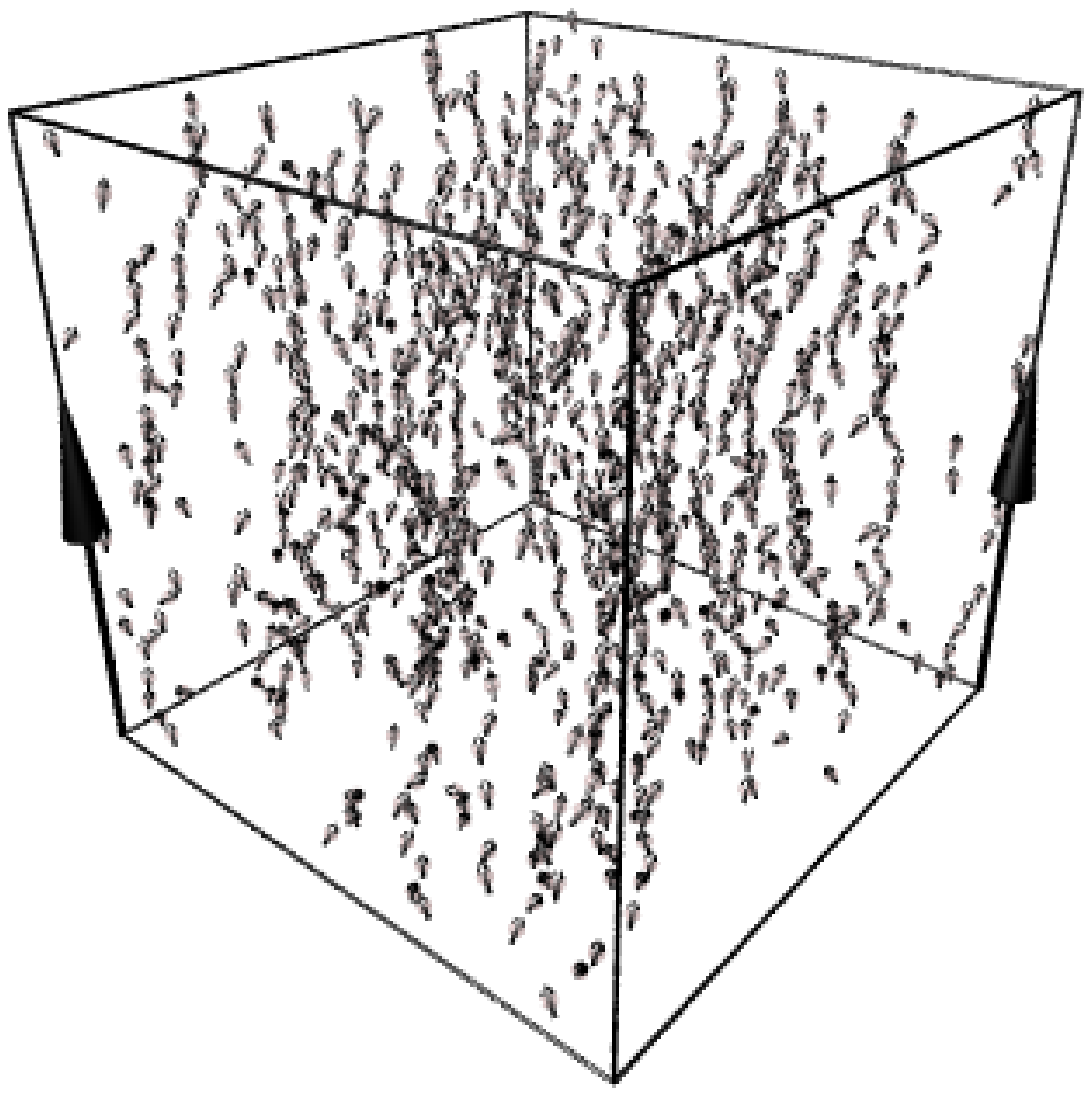}
\end{center}
\caption{Snapshot of the three-dimensional configurations formed
in the system with $\phi=0.0392$ and $\lambda=4$ [studied in Fig.3(a)]
under the magnetic field of $\alpha=0$ (a), $1.0$ (b), $3.0$ (c) and
$9.0$ (d), respectively.
The arrows on the frame indicate the direction of the magnetic field.}
\label{fig8} 
\end{figure}
\newpage
\vspace{4cm}
\begin{figure}[htb]
\begin{center}
\includegraphics*[height=10.0cm]{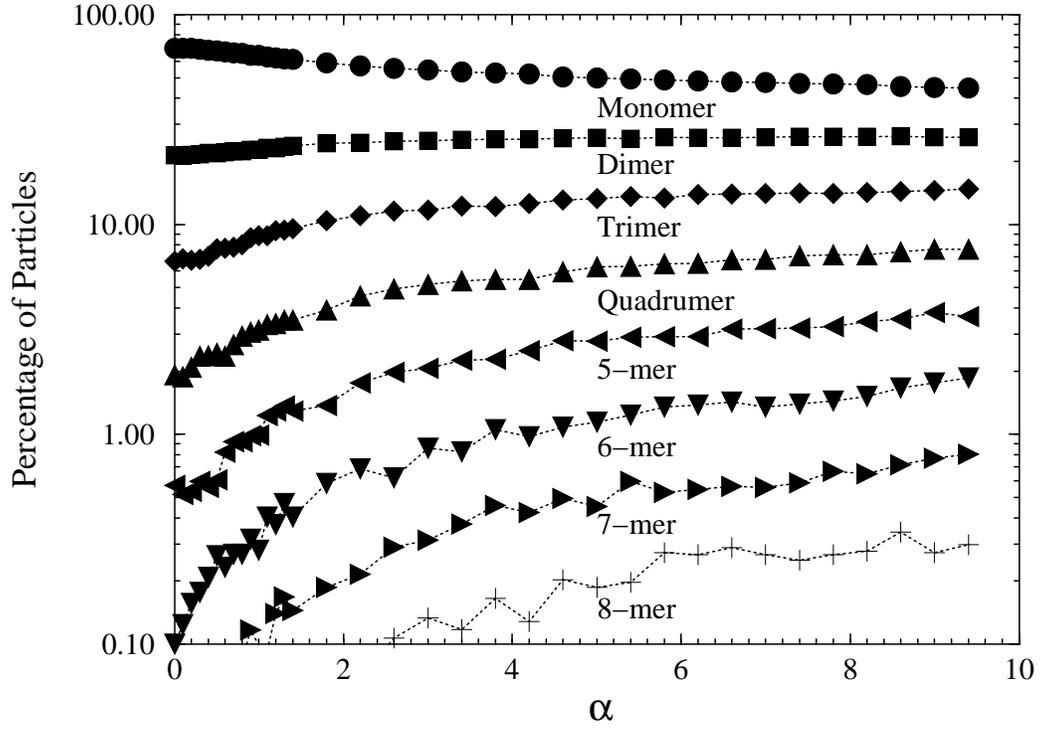}
\end{center}
\caption{Field-dependence of the average percentage of particles
  in $n$-mers for the system with $\phi=0.0392$ and $\lambda=4$.}
\label{fig9}
\end{figure}
\newpage
\vspace{4cm}
\begin{figure}[htb]
\begin{center}
\includegraphics*[height=10.0cm]{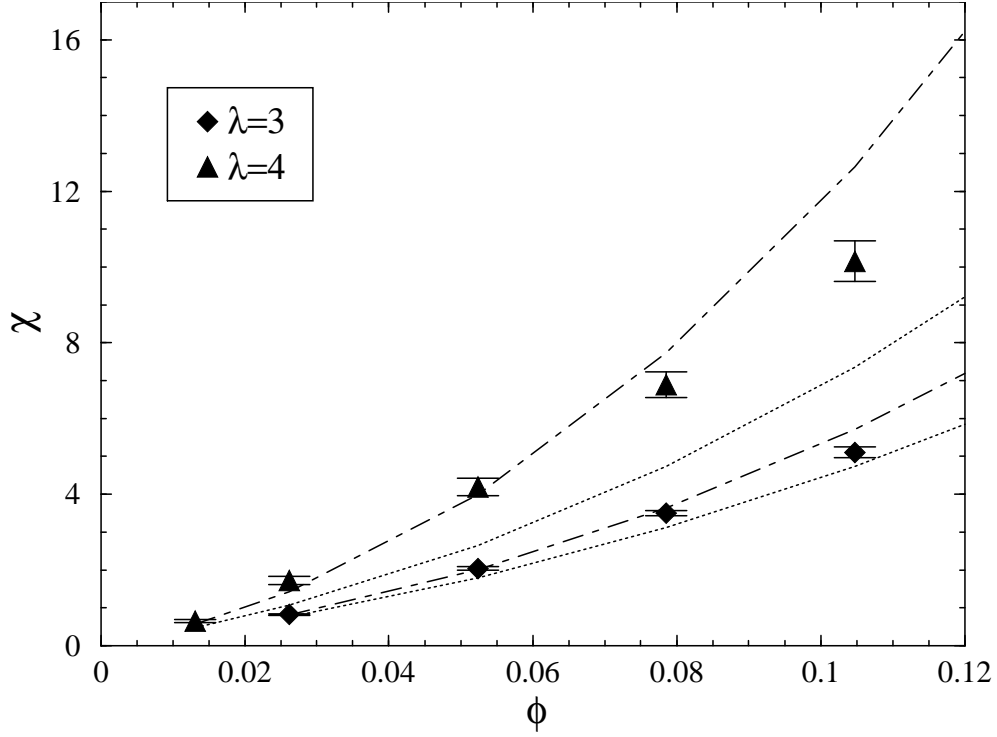}
\end{center}
\caption{Comparison of the simulation results on the initial susceptibility
  with the modified theoretical model. The curves are the theoretical
  prediction of Eq.(\ref{sus3ord}) (dotted) and the modified theoretical
  results obtained by replacing $\chi_L$ with the effective Langevin
  susceptibility $\chi_L^{eff}$ in Eq.(\ref{sus3ord})(dotted-dashed),
  respectively.}
\label{fig10} 
\end{figure}
\newpage
\vspace{3cm}
\begin{figure}[htb]
\begin{center}
  \includegraphics*[height=8.2cm]{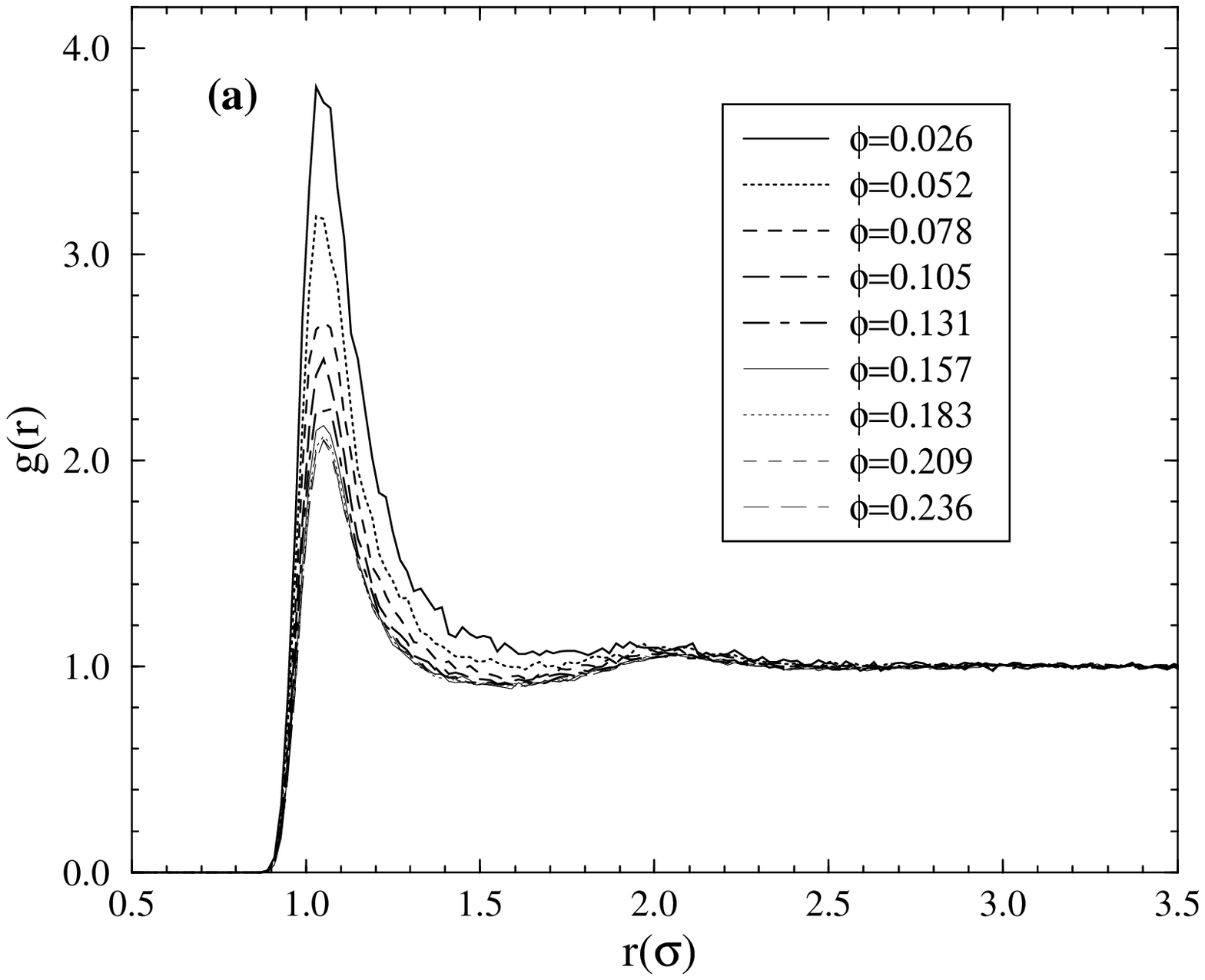}
  \includegraphics*[height=8.2cm]{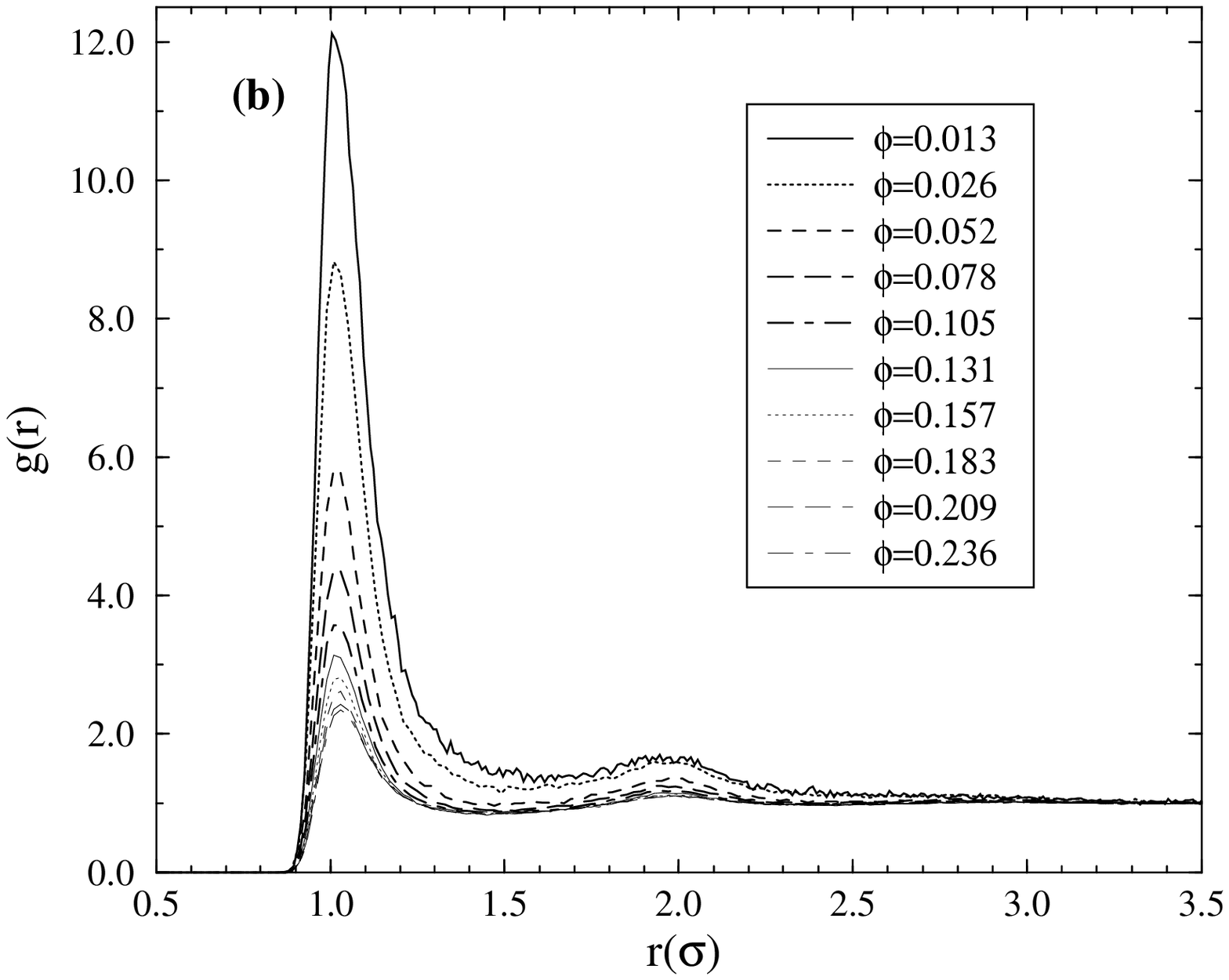}
\end{center}
\caption{Radial distribution function $g(r)$ for the cases
  of $\lambda=3$ (a) and $4$ (b) as a function of $\phi$ at zero-field.}
\label{fig11} 
\end{figure}
\newpage
\vspace{4cm}
\begin{figure}[htb]
\begin{center}
\includegraphics*[height=10.0cm]{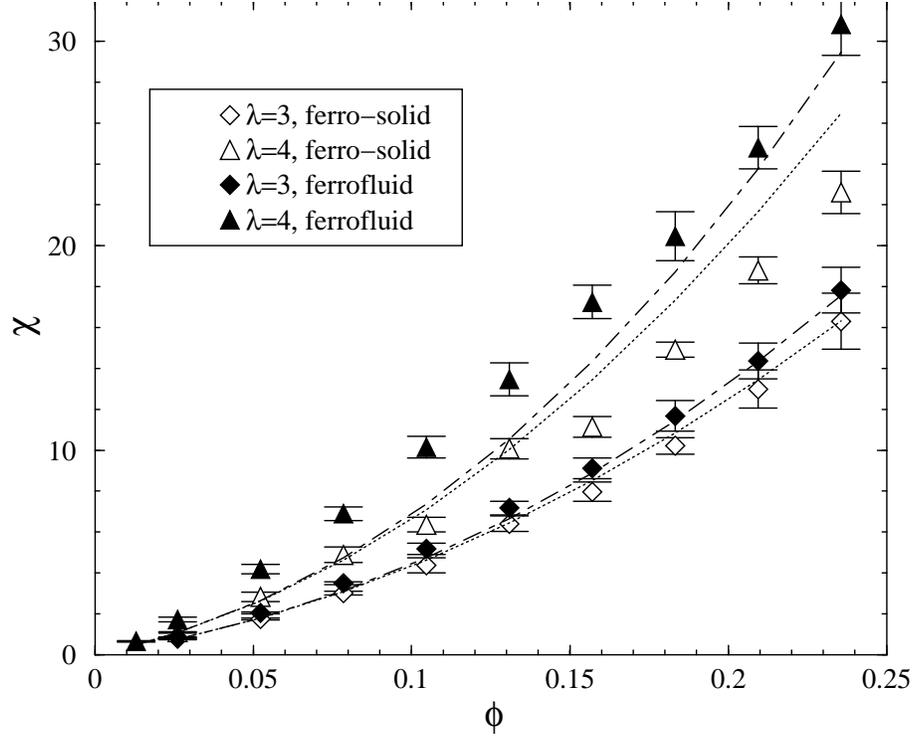}
\end{center}
\caption{Simulation results on the initial susceptibility
  of the ferro-solid systems with $\lambda=3$ and $4$ as a function of the
  volume fraction $\phi$ and the dipolar coupling parameter $\lambda$.  The
  analytical results on the ferrofluid system from
  Eq.(\ref{sus2ord}) (dotted) and Eq.(\ref{sus3ord}) (dotted-dashed),as well
  as the simulation results from Fig.~2 are also included for comparison.  }
\label{fig12} 
\end{figure}

\end{document}